\documentclass[preprintnumbers,amsmath,11pt,amssymb,floatfix,superscriptaddress,nofootinbib]{article}

\def\mytitle#1{\setcounter{equation}{0}
	\setcounter{footnote}{0}
	\begin{flushleft}\Large\textbf{#1}\end{flushleft}
	\vspace{0.25cm}}
\def\myname#1{\leftline{{\large #1}}\vspace{-0.13cm}}
\def\myplace#1#2{\small\begin{flushleft}\textit{#1}\\
		\texttt{#2}\end{flushleft}}

\usepackage{graphicx}
\usepackage{dcolumn}
\usepackage{color}
\usepackage{bm}
\usepackage{amsmath}
\usepackage{amsfonts}
\usepackage{amssymb}

\input epsf

\begin{document}

\mytitle{Universal thermodynamics in different gravity theories: Conditions for generalized second law of thermodynamics and thermodynamical equilibrium on the horizons}
\myname{Saugata Mitra\footnote{saugatamitra20@gmail.com}}
\myplace{Department of Mathematics, Jadavpur University, Kolkata 700032, West Bengal, India.}
\author{Subhajit Saha\footnote {subhajit1729@gmail.com}}
\myplace{Department of Mathematics, Jadavpur University, Kolkata 700032, West Bengal, India.}
\author{Subenoy Chakraborty\footnote {schakraborty.math@gmail.com}}
\myplace{Department of Mathematics, Jadavpur University, Kolkata 700032, West Bengal, India.}{}

\begin{abstract}

The present work deals with a detailed study of universal thermodynamics in different modified gravity theories. The validity of the generalized second law of thermodynamics (GSLT) and thermodynamical equilibrium (TE) of the Universe bounded by a horizon (apparent/event) in f(R)-gravity, Einstein-Gauss-Bonnet gravity, RS-II brane scenario and DGP brane model has been investigated. In the perspective of recent observational evidences, the matter in the Universe is chosen as interacting holographic dark energy model. The entropy on the horizons are evaluated from the validity of the unified first law and as a result there is a correction (in integral form) to the usual Bekenstein entropy. The other thermodynamical parameter namely temperature on the horizon is chosen  as the recently introduced corrected Hawking temperature. The above thermodynamical analysis is done for homogeneous and isotropic flat FLRW model of the Universe. The restrictions for the validity of GSLT and the TE are presented in tabular form for each gravity theory. Finally, due to complicated expressions, the validity of GSLT and TE are also examined from graphical representation, using three Planck data sets.
\end{abstract}

\section{Introduction}

In recent years, the central theme of cosmology has been the interesting fact that our Universe is undergoing an accelerating expansion \cite{r1}. This challenging current problem in cosmology represents a new imbalance in the governing Friedmann equations. Historically, physicists have addressed such imbalances by either introducing new sources, or by altering the governing equations. The standard model of cosmology addresses this imbalance by introducing a new source, named dark energy (an "exotic" matter which has "exotic" properties such as an effective negative pressure) in the Friedmann equations. But the nature of dark energy is completely unknown to us and is an unresolved problem in modern cosmology \cite{r2,r3}. On the contrary, a group of physicists have explored the second route, i.e., a modified gravity approach, on the assumption that at large scales, Einstein's theory of general relativity breaks down and a more general action describes the gravitational field. Such modified theories include f(R)-gravity, Scalar tensor gravity, Einstein-Gauss-Bonnet (EGB) gravity, Brane world gravity and many other  \cite{r4,r5,r6,r7,r8,r9,r10,r11,r12}. The EGB gravity and the Brane world scenario are such models related to gravity theory in higher dimension.
In particular, the Brane world mechanism, whereby matter is confined to the brane while gravity propagates in the bulk, means that extra dimensions can be much longer than in the conventional Kaluza-Klein mechanism, where matter and gravity both propagates in all dimensions.
 In Randall and Sundrum type II (RSII) brane model \cite{r11}, our Universe is a positive tension 3-brane embedded in a five dimensional AdS bulk space-time. The standard model fields are confined on the brane while gravity can propagate in the bulk also. So the effective gravity on the brane is different from the standard Einstein gravity due to the existence of extra dimension. Another simple and well studied model of brane gravity is the Dvali-Gabadadze-Porrati (DGP) brane world model \cite{r12}. In contrast to RSII model, where the extra dimension is of finite size, in DGP brane world model our four dimensional world (3-brane) is embedded in a space-time with an infinite size extra dimension, with the motivation of resolving the cosmological constant problems in super-symmetry breaking \cite{r12}.
These modified theories \cite{r4,r5,r6,r7,r8,r9,r10,r11,r12} are considered as gravitational alternatives for dark energy (DE) and may serve as dark matter (DM) \cite{r13}.\\

Thermodynamical viewpoint of modified gravity theory is an interesting issue in modern theoretical physics. From AdS/CFT correspondence \cite{r14} and black hole thermodynamics \cite{r15}, it has been established that there is a deep connection between gravity and thermodynamics.
Hayward studied thermodynamics for a dynamical black hole \cite{r16,r17} and he introduced the notion of trapping horizon in 4D Einstein gravity for non-stationary spherically symmetric space-times. Subsequently, the idea of Hayward was extended to universal thermodynamics by R. G. Cai and his collaborators \cite{r18,r19,r20,r21}. It has been shown \cite{r22} that Einstein's equations are equivalent to the unified first law. However, they have shown \cite{r18,r19} that the above equivalence in different gravity theories (namely in f(R) and  RSII brane ) is possible by introducing entropy production term. Recently, a modified Bekenstein entropy \cite{r23} was obtained from the validity of the unified first law in different gravity theories by projecting it along any tangential direction to the horizon. The interesting feature of this modified entropy is that the leading order term is the usual Bekenstein entropy and it reduces to Bekenstein entropy in Einstein gravity. In the present work, we shall use this modified entropy of the horizon in different gravity theories namely f(R) gravity, Einstein-Gauss-Bonnet gravity, RSII brane scenario and DGP brane world and examine the validity of the generalized second law of thermodynamics (GSLT) and thermodynamical equilibrium (TE) for Universe bounded by apparent / event horizon. In all the gravity theories the matter in the Universe is chosen as interacting holographic dark energy (HDE).

The paper is organized as follows: section II deals with basic equations for universal thermodynamics while in section III, thermodynamical study in different gravity theories are analyzed. Finally, a brief discussion about the results obtained are presented in section IV.

\section{Universal Thermodynamics: Basic Equations}

In recent years universal thermodynamics has been studied extensively, mostly with an apparent horizon. However, a comparative study of event and apparent horizon from the point of view of the validity of thermodynamical laws has been done by Wang et al. \cite{a1}. Using dark energy fluids, they claimed that universe bounded by an apparent horizon is a Bekenstein system while event horizon is nonphysical in the context of thermodynamics. Subsequently, it has been shown \cite{r31,b1} that assuming first law of thermodynamics, it is possible to have generalized second law of thermodynamics(GSLT) (in any gravity theory) on the event horizon with some realistic conditions. Recently, a correct form of Hawking temperature has been proposed \cite{r33,a2} and as a result, it is possible to obtain the validity of both the thermodynamical laws for any fluid.

In the context of isolated macroscopic systems, the entropy should never decrease, because such a system always evolves towards thermodynamic equilibrium, a state having maximum entropy. Thus for a matter filled universe, bounded by a horizon, the generalized second law of thermodynamics and thermodynamical equilibrium are expressed by the following inequalities \cite{r29,r30},
\begin{eqnarray}\label{1}
\dot{S}_h+\dot{S}_{fh} &\geq & 0
\nonumber
\\
\ddot{S}_h+\ddot{S}_{fh} & < & 0
\end{eqnarray}
where $S_h$ and $S_{fh}$ are the entropies of the horizon and that of the fluid within it respectively. Usually, Clausius relation
\begin{equation}\label{2}
T_hdS_h=dQ_h=-dE_h
\end{equation}
and Gibb's relation \cite{r31,b1,r32}
\begin{equation}\label{3}
T_fdS_{fh}=dE_f+pdV_h
\end{equation}
are used to obtain the entropy variation of the horizon and that of the fluid respectively. In the above ($T_h,~T_f$) stand for the temperature of the horizon and fluid respectively (usually they are chosen identical), $E_h$, $E_f~(=\rho V_h)$ denote respectively the energy flow across the horizon and the total energy of the fluid and $V_h$ is the volume of the fluid bounded by the horizon.

Moreover, in analogy to a dynamical black hole (BH) \cite{r16,r17}, the concept of trapping horizon has been introduced in 4D Einstein gravity for non-stationary spherically symmetric space-times and Einstein equations have been shown to be equivalent to the unified first law (UFL). Then, projecting UFL along the tangential ($\xi$) to the trapping horizon \cite{r18,r19,r20,r21}, one can recover the first law of thermodynamics and the Clausius relation takes the form
\begin{equation}\label{4}
\langle A \psi,\xi \rangle = \frac{\kappa}{8\pi G} \langle dA, \xi \rangle,
\end{equation}
where $\kappa$ is the surface gravity (discussed later) of the horizon, A is the surface area of the horizon and $\psi$ is the usual energy supply vector.

Further, in the context of universal thermodynamics, our Universe should be a non-stationary gravitational system while from the cosmological view point it should be homogeneous and isotropic. So the natural choice is the FLRW Universe- a dynamical spherically symmetric space-time, having only inner trapping horizon (the apparent horizon). In 2005, Cai and Kim \cite{r22} initiated such studies with Hawking temperature ($T_H$) and Bekenstein entropy ($S_B$) on the apparent horizon as
\begin{equation}\label{5}
T_H=\frac{1}{2\pi R_A},~~ S_B=\frac{\pi R_A^2}{G},
\end{equation}
with $R_A$ as the radius of the apparent horizon. They showed the equivalence between (modified) Einstein equations and UFL not only in Einstein gravity but also in Einstein-Gauss-Bonnet gravity and Lovelock gravity. Subsequently, Cai and others \cite{r18,r19,r20,r21} examined the UFL in other modified gravity theories namely, scalar tensor theory \cite{r18} and brane world scenario \cite{r20,r21}. However, one must add entropy production term \cite{a3} to have the Clausius relation in f(R)-gravity theory. Then f(R)-gravity (generalized f(R)) has been studied with a modified version of the horizon entropy. Recently, we have modified the horizon entropy \cite{r23} suitably so that Clausius relation is automatically satisfied. In this context, the present work is an extension in different gravity theories.

The line element for homogeneous and isotropic FLRW model is given by
\begin{eqnarray}\label{6}
ds^2&=&-dt^2+\frac{a^2(t)}{1-kr^2}dr^2+R^2d\Omega_2^2
\nonumber
\\
&=& h_{ab}dx^adx^b+R^2d\Omega_2^2
\end{eqnarray}
where k=0, $\pm 1$ is the curvature scalar, $R=ar$ is the area radius and $h_{ab}=diag(-1,\frac{a^2}{1-kr^2})$ is the metric on 2-space ($x^0=t,~x^1=r$). Using double null co-ordinates $(\xi^\pm)$ the above line element takes the form \cite{r18}:
\begin{equation}\label{7}
ds^2=-2d\xi^+ d\xi^- +R^2 d\Omega_2^2
\end{equation}
with
\begin{equation}\label{8}
\partial_\pm =\frac{\partial}{\partial \xi^\pm}=-\sqrt{2}\left(\frac{\partial}{\partial_t} \mp \frac{\sqrt{1-kr^2}}{a}\frac{\partial}{\partial_r} \right)
\end{equation}
as the future pointing null vectors.

According to Hayward, the trapping horizon ($R_T$) is defined as, $$\partial_+ R|_{R=R_T}=0,$$ i.e.,
\begin{equation}\label{9}
R_T=\frac{1}{\sqrt{H^2+\frac{k}{a^2}}}=R_A.
\end{equation}
The surface gravity is defined as
\begin{equation}\label{10}
\kappa_h=\frac{1}{2\sqrt{-h}}\partial_a(\sqrt{-h}h^{ab}\partial_b R_b),
\end{equation}
which for the above model has the explicit form
\begin{equation}\label{11}
\kappa_h=-\left(\frac{R_h}{R_A}\right)^2 \left(\frac{1-\epsilon}{R_h}\right)
\end{equation}
with $\epsilon =\frac{\dot{R}_A}{2HR_A}$.

Usually in modified gravity theories, the modified Friedmann equations are written as
\begin{equation}\label{12}
H^2+\frac{k}{a^2}=\frac{8\pi G}{3} \rho_t ,
\end{equation}
and
\begin{equation}\label{13}
\dot{H}-\frac{k}{a^2}=-4\pi G (\rho_t + p_t).
\end{equation}
The above equations are nothing but Friedmann equations for non interacting two fluid system- one is the usual fluid of energy density  $\rho$ and thermodynamic pressure $p$, while the second one is termed as effective quantities due to curvature (or other) contributions and we have
\begin{equation}\label{14}
\rho_t=\rho +\rho_e, p_t=p+p_e.
\end{equation}
Usually, the energy supply vector $\psi$ and the work density W are defined as \cite{r16,r17,r18,r19,r20,r21}
\begin{equation}\label{15}
\psi_a=T_a^b \partial_b R + W\partial_a R,~~ W=-\frac{1}{2}T^{ab}h_{ab}.
\end{equation}
So in the present context, the explicit form of W and $\psi$ are
\begin{eqnarray}\label{16}
W&=& \frac{1}{2}(\rho_t -p_t)=\frac{1}{2}(\rho -p)+\frac{1}{2}(\rho_e -p_e)
\nonumber
\\
&=&W_m+W_e
\end{eqnarray}
and
\begin{eqnarray}\label{17}
\psi &=&\psi_m +\psi_e
\nonumber
\\
&=& \lbrace -\frac{1}{2}(\rho +p)HR dt +\frac{1}{2}(\rho +p)a dr \rbrace +\lbrace -\frac{1}{2}(\rho_e +p_e)HRdt+\frac{1}{2}(\rho_e +p_e)a dr \rbrace .
\end{eqnarray}
Note that the heat flow $\delta Q$ in the Clausius relation can be obtained only from the pure matter energy supply $A\psi_m$ by projecting on the horizon. Also the (0, 0)- component of the modified Einstein equations (equation (\ref{12})) is equivalent to the UFL \cite{r16},
\begin{equation}\label{18}
dE=A\psi + WdV,
\end{equation}
where $V=\frac{4}{3}R_h^3$ is the volume bounded by the horizon.

Now for the apparent horizon, the tangent vector $\xi$ can be decomposed in terms of null vectors as
\begin{equation}\label{19}
\xi =\xi_+ \partial_+ +\xi_- \partial_-
\end{equation}
where the ratio of the coefficients can be obtained from the fact that $\partial_+ R_T=0$,
\begin{equation}\label{20}
i.e.,~~\frac{\xi_+}{\xi_-}=-\frac{\partial_-\partial_+ R_T}{\partial_+\partial_+ R_T}
\end{equation}
As for the present model
\begin{equation}\label{21}
\partial_-\partial_+ R_A = \frac{4}{R_A}(1-\epsilon), \partial_+\partial_+ R_A =-\frac{4\epsilon}{R_A},
\end{equation}
so in (t, r)-coordinate, $\xi$ can have the explicit form
\begin{equation}\label{22}
\xi=\frac{\partial}{\partial t} - (1-2\epsilon)Hr\frac{\partial}{\partial t}.
\end{equation}
Thus projecting the UFL along $\xi$, the first law of thermodynamics takes the form \cite{r18,r19,r20,r21},
\begin{equation}\label{23}
\langle dE, \xi \rangle =\frac{\kappa}{8\pi G} \langle dA, \xi \rangle +\langle WdV, \xi \rangle ,
\end{equation}
and consequently,
\begin{equation}\label{24}
\delta Q=\langle A\psi_m , \xi \rangle = \frac{\kappa_A}{8\pi G} \langle dA, \xi \rangle -\langle A\psi_e, \xi \rangle ,
\end{equation}
or in the explicit form
\begin{equation}\label{25}
\delta Q=-\frac{2 \epsilon (1-\epsilon)}{G}HR_A +A(1-\epsilon)HR_A(\rho_e +p_e).
\end{equation}
Further, using Hawking temperature on the apparent horizon as
\begin{equation}\label{26}
T_A=\frac{\kappa_A}{2\pi}=\frac{1-\epsilon}{2\pi R_A},
\end{equation}
and using the Clausius relation: $\delta Q= TdS$, the entropy on the apparent horizon can be taken to be
\begin{equation}\label{27}
dS_A=\frac{2\pi R_A dR_A}{G}- 8 \pi ^2 H R_A^4 (\rho_e +p_e) dt
\end{equation}
i.e.,
\begin{equation}\label{28}
S_A=\frac{A_A}{4G}-8 \pi^2 \int HR_A^4 (\rho_e +p_e)dt,
\end{equation}
which shows that the entropy on the apparent horizon differs from the usual Bekenstein entropy by a correction term (in integral form).

On the other hand, for the event horizon, as $d\xi^\pm=dt \mp adr$ is the one form orthogonal to the surface of the event horizon, so the tangent vector $\xi$ can be taken as \cite{r23}
\begin{equation}\label{29}
\xi = \frac{\partial}{\partial t}-\frac{1}{a}\frac{\partial}{\partial r},
\end{equation}
and consequently the modified entropy on the event horizon has the explicit form
\begin{equation}\label{30}
S_E=\frac{A_E}{4G}-4\pi^2 \int \left(\frac{R_A^2 R_E}{1-\epsilon}\right)\left(\frac{HR_E+1}{HR_E-1}\right) (\rho_e +p_e)dR_E.
\end{equation}
which also has the Bekenstein entropy as the leading term.

Before ending this section, it is worthy to mention that in the cosmological context, one can consider the following entropy bounds:\\
(i) the Bekenstein bound,\\
(ii) the holographic Bekenstein-Hawking bound \cite{a6}\\
and (iii) Cardy-Verlinde (C-V) bound \cite{a5}.\\
The Bekenstein bound is supposed to hold for systems with limited self-gravity. In cosmological perspective, this implies that the Hubble radius ($H^{-1}$) is larger than the radius (R) of the Universe. On the other hand, in a strongly self-gravitating universe (i.e., HR $\geq$ 1), one has to take into account the possibility of BH formation. Thus using the general philosophy of the holographic principle, one gets $S_B$ (Bekenstein entropy) $\leq S_{BH}$ for weak self-gravity while $S_B \geq S_{BH}$ for strong self-gravity, where $S_{BH}=(n-1)\frac{V}{4GR}$, is the holographic Bekenstein-Hawking entropy of a black hole of same size as the universe (n+1 is the dimension of the space-time). One may note that $S_{BH}$ grows like an area instead of the volume and for a closed Universe, it is the closest one to the usual expression $\frac{A}{4G}$. The limiting situation i.e, $S_B=S_{BH}$ when $R=\frac{1}{H}$ is termed as Hubble bound. Finally, the C-V bound is valid through out the cosmological evolution and is associated with the Casimir energy. Also it has been showed that \cite{a6} Friedmann equation in radiation dominated FRW Universe can be written in an analogous form of C-V formula, an entropy formula for a conformal field theory.

\section{Different Gravity Theories and Thermodynamical Analysis}

In this section we shall examine the validity of the generalized second law of thermodynamics and the thermodynamical equilibrium for the following modified gravity theories:\\
i) f(R)-gravity,\\
ii) Einstein-Gauss-Bonnet gravity,\\
iii)RSII brane scenario\\
and iv) DGP brane world.

Let us consider our Universe to be homogeneous and isotropic flat FRW model and the line element is given by (for simplicity $8\pi =1=G$)
\begin{center}
$ds^2=-dt^2+a(t)^2[dr^2+r^2d\Omega_2^2]$.
\end{center}
where $a(t)$ is the scale factor of the Universe and $d\Omega_2^2$ is the metric on the unit 2-space.\\

Regarding matter distribution, we assume that Universe is filled with holographic dark energy (HDE) interacting with dark matter (DM) in the form of dust. The argument behind choice of interaction models is that they are favoured by observed data obtained from the Cosmic Microwave Background (CMB) \cite{r24} and matter distribution at large scales \cite{r25}. Further, Das et al. \cite{r26} and Amendola et al. \cite{r27} showed that an interaction (between HDE and DM in the dust form) model of the Universe mimics the observationally measured phantom equation of state as compared to noninteracting models, which may predict a non-phantom type of  equation of state.
The variable equation of state parameter for the HDE has the form \cite{r28},
\begin{equation}
\omega_d=-\frac{1}{3}-\frac{2\sqrt{\Omega_d}}{3c}-\frac{b^2}{\Omega_d},
\end{equation}

where c is a dimensionless parameter (estimated from observations), and the interaction term has the form $3b^2H(\rho_d+\rho_m)$, where $b^2$ is the coupling parameter between DE and DM and $\rho_m,~\rho_d$ are the energy densities of the two dark components namely dark matter and dark energy. The density parameter $\Omega_d$ is given by
\begin{equation}
\Omega_d'=\Omega_d \left[(1-\Omega_d)\left(1+\frac{2\sqrt{\Omega_d}}{c}\right)-3b^2\right]
\end{equation}
where $'=\frac{\partial}{\partial x},~ x=lna$.\\
The velocities of the apparent ($v_A$) and event horizon ($v_E$) can be expressed as
\begin{equation}
v_A=\frac{3}{2}\left[(1-b^2)-\frac{\Omega_d}{3}\left(1+\frac{2\sqrt{\Omega_d}}{c}\right)\right]
\end{equation}
and
\begin{equation}
v_E=\left(\frac{c}{\sqrt{\Omega_d}}-1\right)
\end{equation}

\subsection{$f(R)$-gravity}

In $f(R)$ gravity, the modified Einstein-Hilbert action can be written as (in Jordan frame) \cite{r4}
\begin{equation}
A=\frac{1}{2} \int d^4x \sqrt{-g}f(R)+A_m,
\end{equation}
with $A_m$ as the matter action (here we have considered $8\pi=1$ and G=1 ). Now, variation of $A$ with respect to the metric tensor $g_{\mu \nu}$ gives the modified field equations in $f(R)$ gravity as
\begin{equation}
R_{\mu \nu} \frac{\partial f}{\partial R}-\frac{1}{2}g_{\mu \nu}f(R)-\nabla_\mu \nabla_\nu\left(\frac{\partial f}{\partial R}\right)+g_{\mu \nu}\nabla^2\left(\frac{\partial f}{\partial R}\right)=T_{\mu \nu},
\end{equation}
where $T_{\mu}^\nu = diag(-\rho , p, p, p)$ is the energy-momentum tensor for the matter field in the form of perfect fluid with $\rho=\rho_m+\rho_d,~p=p_d$. Here ($\rho_d, p_d$) are the energy density and thermodynamic pressure of the HDE while $\rho_m$ is the energy density of the dark matter. In the present thermodynamical analysis chameleon scenario \cite{c1} has not been considered explicitly. However, instead of using HDE, one may use chameleon scalar field to explain the recent observations and this will not affect the thermodynamics to a great extend.

In particular, for a viable $f(R)$-gravity theory if we take
\begin{equation}
f(R)=R+F(R)
\end{equation}
then the explicit form of the modified field equations for FRW metric are given by
\begin{equation}
H^2=\frac{1}{3}\rho_t
\end{equation}
and
\begin{equation}
\dot{H}=-\frac{1}{2}(\rho_t+p_t),
\end{equation}

The effective energy density $\rho_e$ and effective pressure $p_e$ due to the curvature contribution have the expressions
\begin{equation}
\rho_e=\left[-\frac{1}{2}(F-RF_1)-3H\dot{F_1}-3F_1H^2\right]
\end{equation}
and
\begin{equation}
\rho_e+p_e=\left[\ddot{F_1}-H\dot{F_1}-2H^2v_AF_1\right],
\end{equation}
where $F_1=\frac{dF}{dR}$, $R=6(\dot{H}+2H^2)$ is the Ricci scalar.\\

The energy conservation relations are
\begin{center}
$\dot{\rho}_t+3H(\rho_t+p_t)=0~,~~\dot{\rho}+3H(\rho+p)=0$.
\end{center}
So we have $$\dot{\rho}_e+3H(\rho_e+p_e)=0.$$
Now using equations (\ref{28}) and (\ref{30}), the expression of entropy for apparent horizon ($S_A$) and event horizon ($S_E$) are given by,
\begin{equation}
S_A=\frac{A_A}{4}-\frac{1}{8}\int \left(\ddot{F_1}-H\dot{F_1}+2F_1\dot{H}\right)HR_A^4dt
\end{equation}
and
\begin{equation}
S_E=\frac{A_E}{4}-\frac{1}{16}\int \left(\frac{R_A^2 R_E}{1-\epsilon}\right)\frac{HR_E+1}{HR_E-1}\left(\ddot{F_1}-H\dot{F_1}-\frac{4F_1\epsilon}{R_A^2}\right)dR_E.
\end{equation}
Here $R_A$ and $R_E$ are the radii of the apparent horizon (AH) and the event horizon (EH) respectively.\\

Thus using equation (\ref{3}) for the fluid entropy, the first and second derivatives of the total entropy can be written as,
\begin{equation}\label{gsfa}
\dot{S}_{TA}=\frac{R_Av_A}{4}-\frac{R_A^3}{8}\left(\ddot{F_1}-H\dot{F_1}-2H^2v_AF_1+\frac{v_A}{6(2-v_A)}\{\rho-\frac{F-RF_1}{2}-3H\dot{F_1}-3H^2F_1\}\right),
\end{equation}

\begin{equation}\label{gsfe}
\dot{S}_{TE}=\frac{R_Ev_E}{4}-\frac{R_A^2R_E}{2(2-v_A)}\left(\frac{v_E+2}{4}(\ddot{F_1}-H\dot{F_1}-2H^2v_AF_1)+\frac{v_A}{3}\{\rho-\frac{F-RF_1}{2}-3H\dot{F_1}-3H^2F_1\}\right),
\end{equation}
$$\ddot{S}_{TA}=\frac{R_Af_A}{4}\{1-\frac{2R_A^2}{3(v_A-2)^2}(v_A^2-4v_A+2)\{\rho-\frac{F-RF_1}{2}-3H\dot{F_1}-3H^2F_1\}\}-\frac{R_A^2}{8} \left(3v_A(\ddot{F_1}-H\dot{F_1}-2H^2v_AF_1)\right.$$
\begin{equation}\label{tefa}
\left.+8R_A(\dddot{F_1}-H\ddot{F_1}-H^2v_A\dot{F_1}+2F_1(2H^3v_A^2-H^2f_A))\right)-\frac{R_A^2v_A^2(v_A-1)}{6(v_A-2)}\{\rho-\frac{F-RF_1}{2}-3H\dot{F_1}-3H^2F_1\},
\end{equation}
$$\ddot{S}_{TE}=\frac{R_Ef_E}{4}\{1-\frac{R_A^2}{2(2-v_A)}(\ddot{F_1}-H\dot{F_1}-2H^2v_AF_1)\}+\frac{v_E^2}{4}-\frac{R_A^2R_E}{8(2-v_A)}\left[(v_E+2)\{\left(2\frac{v_A}{R_A}+\frac{v_E}{R_E}+\frac{f_A}{2-v_A}\right)\right.$$
$$\times (\ddot{F_1}-H\dot{F_1}-2H^2v_AF_1)+(\dddot{F_1}-H\ddot{F_1}-H^2v_A\dot{F_1}+2F_1(2H^3v_A^2-H^2f_A))\}$$
\begin{equation}\label{tefe}
\left.+\frac{4v_A\{\rho-\frac{F-RF_1}{2}-3H\dot{F_1}-3H^2F_1\}}{3}\{\frac{v_E}{R_E}+\frac{2f_A}{v_A(2-v_A)}\}\right]
\end{equation}
where $f_A=\dot{v}_A$ and $f_E=\dot{v}_E$ are the accelerations of the apparent and the event horizon respectively and $S_{TA}=S_A+S_{fA}$ and $S_{TE}=S_E+S_{fE}$ are respectively the total entropy of the system bounded by the apparent and event horizons.

From the above equations (\ref{gsfa})-(\ref{tefe}), the conditions (i.e., inequations (\ref{1})) for the validity of GSLT and TE are shown in Table-I. As the expressions are very complicated, so to make some conclusive remarks we shall plot the expressions for $\dot{S}_{TA}$, $\dot{S}_{TE}$, $\ddot{S}_{TA}$ and $\ddot{S}_{TE}$ against the coupling parameter $b^2$ in figure 1 (a)- (d) for the choice of f(R) as $R+R^2$ \cite{r34}. In the figures, the density parameter $\Omega_d$ and the dimensionless parameter c are estimated from three Planck data sets \cite{r35} (see Table-II).
\newpage
\begin{center}
{\bf Table I}: Condition(s) required for GSLT and TE to hold in $f(R)$-gravity (Jordan frame)
\end{center}
\begin{center}
\begin{tabular}{|p{1.55cm}|p{1.2cm}|p{15cm}|}
\hline \begin{center} GSLT/TE \end{center} & \begin{center} Horizon \end{center} & \begin{center} Condition(s) \end{center}\\
\hline \hline \begin{center} GSLT \end{center} & \begin{center} AH \end{center} &  \begin{center} $v_A(1+F_1) - \frac{1}{H^2}\left[\frac{1}{2}(\ddot{F_1}-H\dot{F_1})+\frac{2}{3}\frac{v_A(v_A-1)}{(v_A-2)}\{\rho-\frac{1}{2}(F-RF_1)-3H\dot{F_1}-3H^2F_1\}\right] \geq 0$ \end{center}\\
\hline \begin{center} GSLT \end{center} & \begin{center} EH \end{center} &  \begin{center} $v_E \lesseqgtr  \frac{\left[\frac{2}{H^2}\{\ddot{F_1}-H\dot{F_1}-2H^2v_AF_1\}+\frac{2v_A}{3}\{\rho-\frac{1}{2}(F-RF_1)-3H\dot{F_1}-3H^2F_1\}\right]}{\{2(2-v_A)-\frac{1}{H^2}(\ddot{F_1}-H\dot{F_1}-2H^2v_AF_1)\}}$\end{center} \begin{center} according as $1 \lessgtr \frac{1}{2H^2(2-v_A)}(\ddot{F_1}-H\dot{F_1}-2H^2v_AF_1)$ \end{center}\\
\hline \begin{center} TE \end{center} & \begin{center} AH \end{center} &  \begin{center} $f_A\left[1-\frac{2(v_A^2-4v_A+2)}{3H^2(v_A-2)^2}\{\rho-\frac{1}{2}(F-RF_1)-3H\dot{F_1}-3F_1H^2\}+8F_1\right]< \frac{v_A}{2H}\left[\dddot{F_1}(\frac{8}{v_AH})+\ddot{F_1}(3-\frac{8}{v_A})+\dot{F_1}(-11H-\frac{4v_A(v_A-1)}{v_A-2})\right. $\end{center} \begin{center}$\left.+F_1 \left(-6H^2v_A+32v_A^2H^2+\frac{4Hv_A(v_A-1)}{v_A-2}+\frac{2Rv_A(v_A-1)}{3(v_A-2)H}\right)+\frac{4v_A(v_A-1)}{3H(v_A-2)}\{\rho-\frac{F}{2}\}\right]$ \end{center}\\
\hline \begin{center} TE \end{center} & \begin{center} EH \end{center} & \begin{center} $H^2f_E\{2(2-v_A)-\frac{\ddot{F_1}-H\dot{F_1}-2H^2v_AF_1}{H^2}\}\lessgtr (v_E+2)\{\left(2v_AH+\frac{v_E}{R_E}+\frac{f_A}{2-v_A}\right)(\ddot{F_1}-H\dot{F_1}-2H^2v_AF_1)+\left(\dddot{F_1}-H\ddot{F_1}-v_AH^2\dot{F_1}+2F_1H^2(2Hv_A^2-f_A)\right)\}$\end{center}\begin{center}$+\frac{4v_A}{3}\{\rho-\frac{F-RF_1}{2}-3H\dot{F_1}-3F_1H^2\}\{\frac{v_E}{R_E}+\frac{2f_A}{v_A(2-v_A)}\}-\frac{2H^2v_E^2}{R_E}(2-v_A)$,\end{center} \begin{center} according as $v_A\lessgtr 2$ \end{center}\\
\hline
\end{tabular}
\end{center}

\begin{center}
{\bf Table II: Planck Data Sets}
\end{center}
\begin{center}
\begin{tabular}{|c|c|c|c|}
\hline Sl. No. & Data Sets & c & $\Omega_d$ \\
\hline \hline 1 & Planck+CMB+SNLS3+lensing & 0.603 & 0.699 \\
\hline 2 & Planck+CMB+Union 2.1+lensing & 0.645 & 0.679 \\
\hline 3 & Planck+CMB+BAO+HST+lensing & 0.495 & 0.745\\
\hline
\end{tabular}
\end{center}
\begin{figure}
\begin{minipage}{0.4\textwidth}
\includegraphics[width=1.0\linewidth]{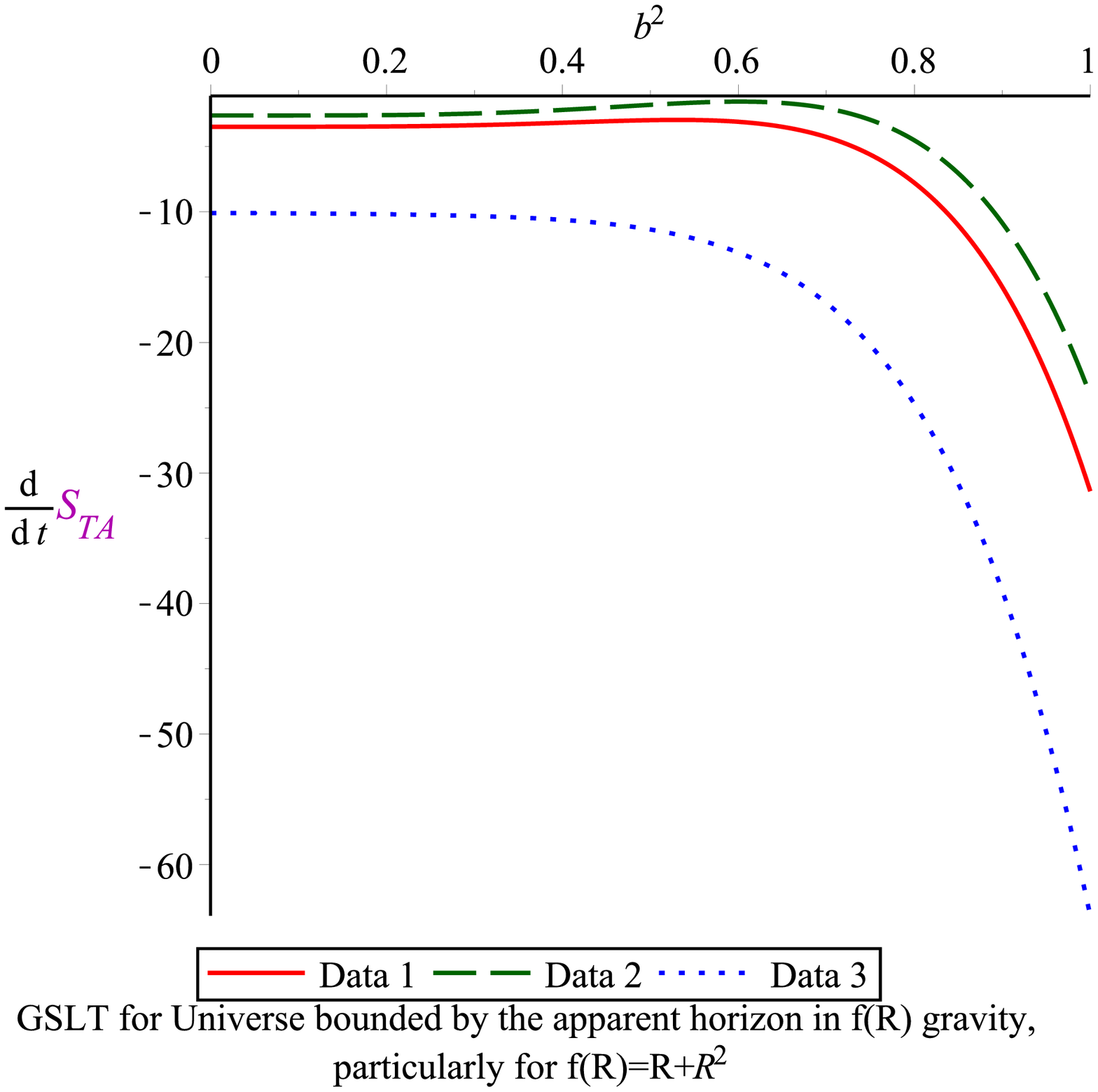}
(a)
\end{minipage}
\begin{minipage}{0.4\textwidth}
\includegraphics[width=1.0\linewidth]{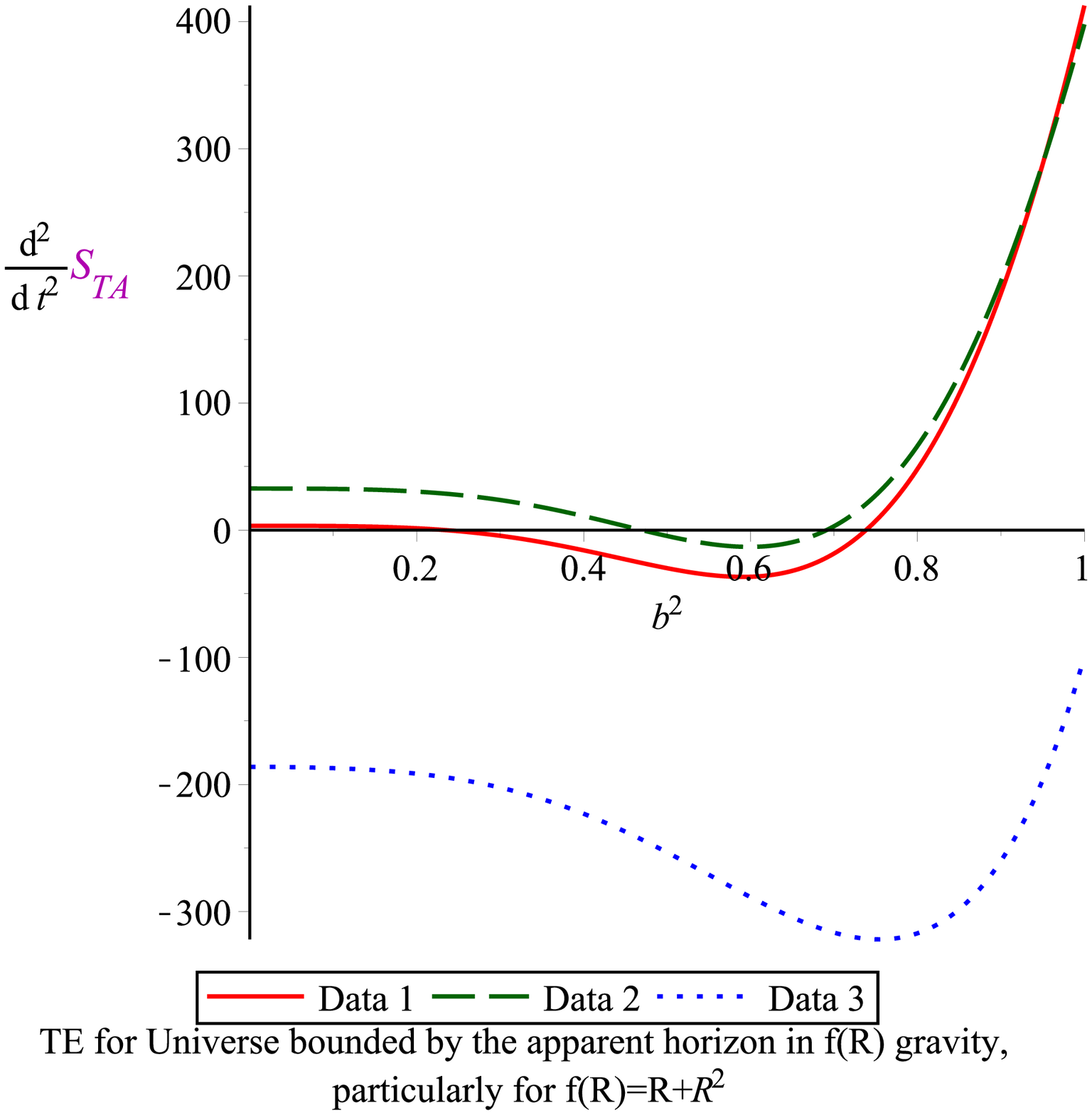}
(b)
\end{minipage}
\begin{minipage}{0.4\textwidth}
\includegraphics[width=1.0\linewidth]{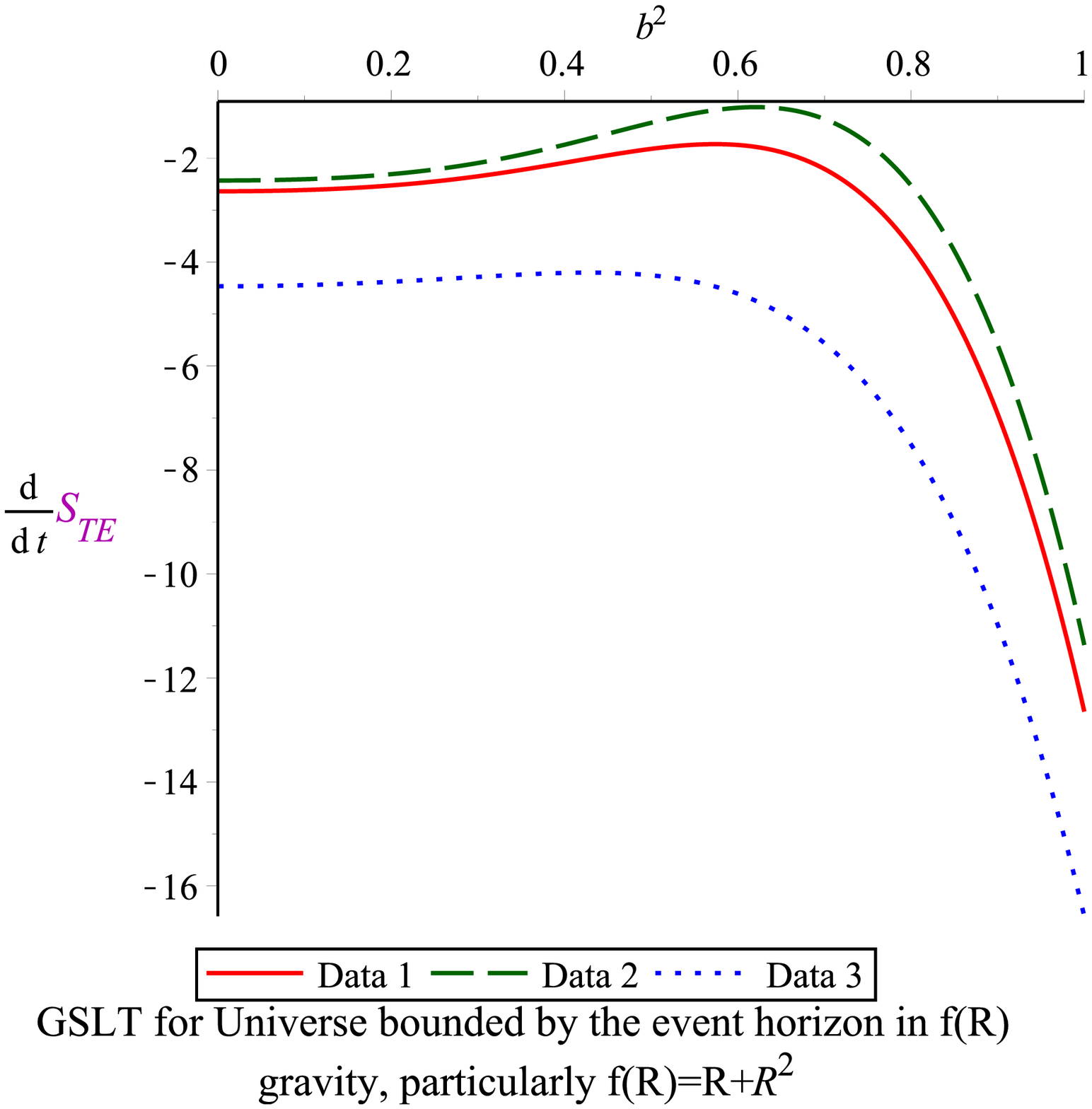}
(c)
\end{minipage}
\begin{minipage}{0.4\textwidth}
\includegraphics[width=1.0\linewidth]{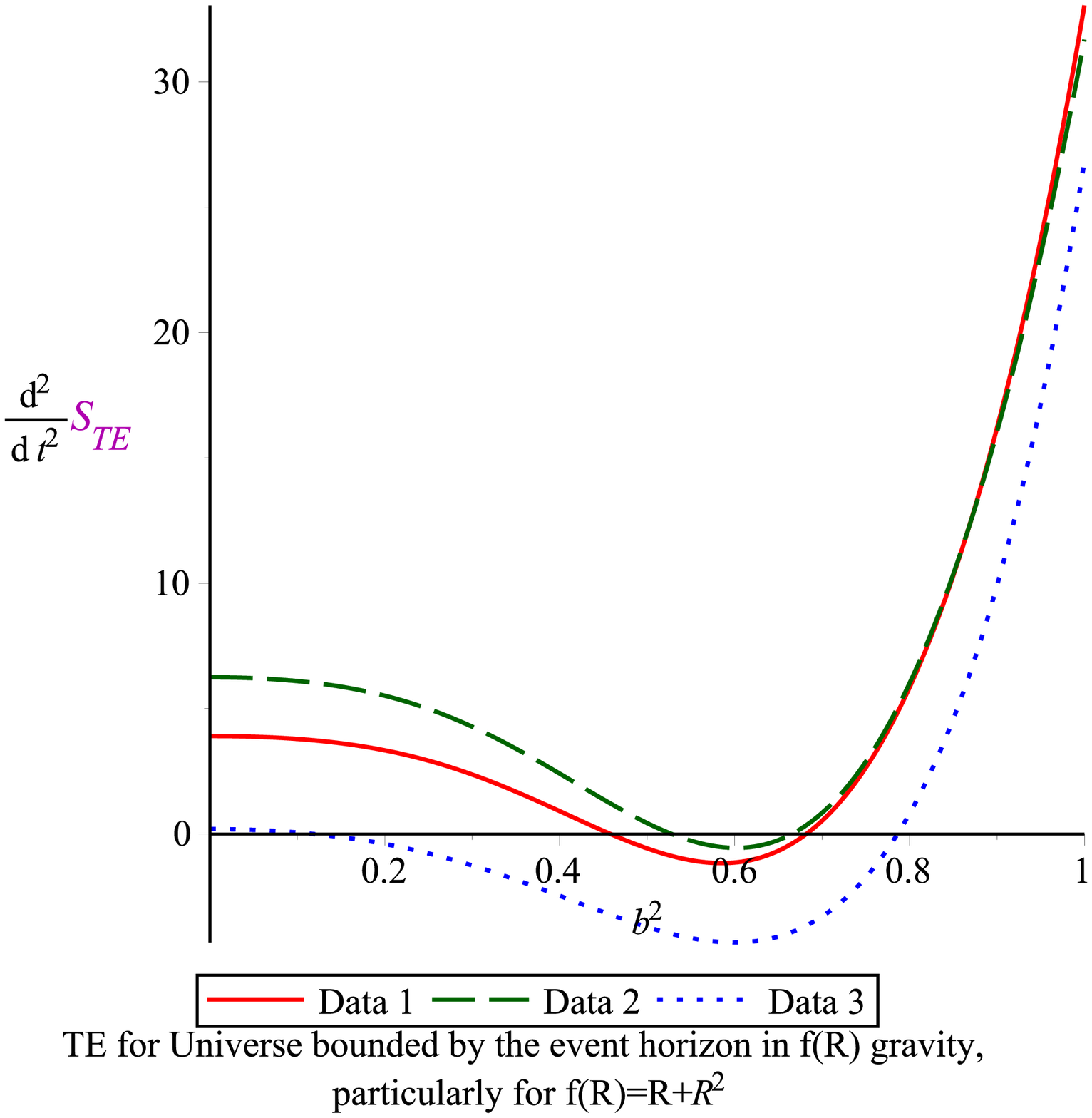}
(d)
\end{minipage}
\caption{The plots show GSLT and TE for Universe bounded by apparent/ event horizon for f(R)-gravity considering f(R)=$R+R^2$}
\end{figure}
Planck data reduces the error by 30\%-60\% when compared to  Wilkinson Microwave Anisotropy Probe (WMAP)-9 data. The accuracy is further increased if the External Astrophysical data sets (EADS) as well as lensing data is taken into account. Common EADS include the Baryonic Acustic Oscillation (BAO) measurements from 6dFGS+SDSS DR7(R) + BOSS DR9, Hubble constant estimated from the Hubble Space Telescope (HST) and supernova data sets SNLS3 and Union 2.1.\\

In the Einstein frame (obtained by considering a conformal transformation $\tilde{g_{ab}}=e^\phi g_{ab}$ where $\phi=ln[1+F_1(R)]$), the Friedmann equations can be written as $[23]$
\begin{eqnarray}
H^2&=&\frac{1}{3}\left(\rho+ \left(\frac{3}{4}\dot{\phi}^2+\frac{1}{2}V(\phi)\right)\right),
\nonumber
\\
\dot{H}&=&-\frac{1}{2}\left(\rho+p+\left(\frac{3}{2}\dot{\phi}^2\right)\right)
\end{eqnarray}
and the conservation equation is
\begin{center}
$\ddot{\phi}+3H\dot{\phi}+\frac{1}{3}\frac{\partial V}{\partial \phi} = 0$,
\end{center}
where $V(\phi)$ is the effective potential.\\

Thus, in the Einstein frame the expressions for $\rho _t$ and $\rho _e+p_e$ become
\begin{eqnarray}\label{ef}
\rho_t &=& \rho+ \frac{3}{4}\dot{\phi}^2+\frac{1}{2}V(\phi),
\nonumber
\\
\rho_e+p_e &=& \left(\frac{3}{2}\dot{\phi}^2\right),
\nonumber
\\
\textrm{and}~\frac{\partial (\rho_e+p_e)}{\partial t}&=&3\dot{\phi}\ddot{\phi}.
\end{eqnarray}
The entropy on the apparent and the event horizon are respectively given by (using equations (\ref{28}) and (\ref{30}))
\begin{center}
 $S_A=\frac{A_A}{4}-\frac{3}{16} \int \left(\dot{\phi}^2HR_A^4\right)dt$
\end{center}
and
\begin{center}
$ S_E=\frac{A_E}{4}-\frac{3}{32} \int \left(\frac{R_A^2R_E}{(1-\epsilon)}\right)\left(\frac{HR_E+1}{HR_E-1}\right)\dot{\phi}^2dR_E$.
\end{center}
Note that the scalar field $\phi$ in the Einstein frame corresponds to a representative form of Ricci scalar in Jordan frame. In our scenario, the
Einstein frame is the physical frame which gives self gravity of the scalar field's effective potential $V(\phi)$.\\

Thus the variation of the total entropy both at the apparent and event horizon are given by
\begin{equation}
\dot{S}_{TA}=\frac{R_A v_A}{4}-\frac{R_A^3}{8}(\rho_e+p_e)+\frac{R_A^3v_A(v_A-1)\rho_t}{6(2-v_A)},
\end{equation}
\begin{equation}
\dot{S}_{TE}=\frac{R_Ev_E}{4}-\frac{R_A^2R_E}{2(2-v_A)}\{\frac{v_E+2}{4}(\rho_e+p_e)+\frac{v_A\rho_t}{3}\},
\end{equation}
\begin{equation}
\ddot{S}_{TA}=\frac{R_Af_A}{4}\{1-\frac{2R_A^2\rho_t}{3(v_A-2)^2}(v_A^2-4v_A+2)\}-\frac{R_A^2}{8}\{3v_A(\rho_e+p_e)+8R_A\frac{\partial (\rho_e+p_e)}{\partial t}\}-\frac{R_A^2v_A^2(v_A-1)\rho_t}{6(v_A-2)},
\end{equation}
$$\ddot{S}_{TE}=\frac{R_Ef_E}{4}\{1-\frac{R_A^2(\rho_e+p_e)}{2(2-v_A)}\}+\frac{v_E^2}{4}-\frac{R_A^2R_E}{8(2-v_A)}\{(v_E+2)\left(2\frac{v_A}{R_A}+\frac{v_E}{R_E}+\frac{f_A}{2-v_A}\right)(\rho_e+p_e)$$
\begin{equation}
+\frac{\partial (\rho_e+p_e)}{\partial t}\}+\frac{4v_A\rho_t}{3}\{\frac{v_E}{R_E}+\frac{2f_A}{v_A(2-v_A)}\}.
\end{equation}
Here $\rho_t$, $(\rho_e+p_e)$, $\frac{\partial (\rho_e+p_e)}{\partial t}$ are to be substituted from equation (\ref{ef}).

In Table-III, we have presented restrictions for the validity of GSLT and TE for both the horizons.
\newpage
\begin{center}
{\bf Table III}: Condition(s) required for GSLT and TE to hold in $f(R)$-gravity (Einstein frame)
\end{center}
\begin{center}
\begin{tabular}{|p{2cm}|p{2cm}|p{12cm}|}
\hline \begin{center} GSLT/TE \end{center} & \begin{center} Horizon \end{center} & \begin{center} Condition(s) \end{center}\\
\hline \hline \begin{center} GSLT \end{center} & \begin{center} AH \end{center} &  \begin{center} $v_A - \frac{3\dot{\phi}^2}{4H^2}+\frac{2v_A(v_A-1)}{3H^2(v_A-2)}\{\rho+\frac{3\dot{\phi}^2}{4}+\frac{V(\phi)}{2}\}\geq 0$ \end{center}\\
\hline \begin{center} GSLT \end{center} & \begin{center} EH \end{center} &  \begin{center} $v_E\lesseqgtr \frac{2\{\left(\frac{3}{2}+\frac{v_A}{2}\right)\dot{\phi}^2+\frac{2v_A\rho}{3}+\frac{v_AV(\phi)}{3}\}}{2H^2(2-v_A)-\frac{3\dot{\phi}^2}{2}}$\end{center}  \begin{center} according as $1\lessgtr \frac{3\dot{\phi}^2}{4H^2(2-v_A)}$ \end{center}\\
\hline \begin{center} TE \end{center} & \begin{center} AH \end{center} &  \begin{center} $f_A\{1-\frac{2(v_A^2-4v_A+2)(\rho+\frac{3\dot{\phi}^2}{4}+\frac{V(\phi)}{2})}{3H^2(v_A-2)^2}\}<\frac{v_A}{2H}\{\dot{\phi}\left(\frac{24\ddot{\phi}}{Hv_A}+\dot{\phi}\left(\frac{9}{2}+\frac{v_A(v_A-1)}{H(v_A-2)}\right)\right)+\frac{4v_A(v_A-1)}{3H(v_A-2)}\{\rho+\frac{V(\phi)}{2}\}\}$ \end{center}\\
\hline \begin{center} TE \end{center} & \begin{center} EH \end{center} & \begin{center} $f_E \{2(2-v_A)-\frac{3\dot{\phi}^2}{2H^2}\}\lessgtr \frac{1}{H^2}\left[(v_E+2)\{\left(2v_AH+\frac{v_E}{R_E}+\frac{f_A}{2-v_A}\right)\frac{3\dot{\phi}^2}{2}+3\ddot{\phi}\dot{\phi}\}\right.$\end{center}\begin{center}$\left.+\frac{4v_A}{3}\{\rho+\frac{3\dot{\phi}^2}{4}+\frac{V(\phi)}{2}\}\{\frac{v_E}{R_E}+\frac{2f_A}{v_A(2-v_A)}\}\right]-\frac{2v_E^2}{R_E}(2-v_A)$,\end{center} \begin{center} according as $v_A\lessgtr 2$ \end{center}\\
\hline
\end{tabular}
\end{center}

 \subsection{Einstein-Gauss-Bonnet (EGB) gravity}

In Einstein-Gauss-Bonnet gravity, the action in (4+1) dimensions can be written as
\begin{center}
$I=\frac{1}{2}\int (\sqrt{-g}(R+\alpha R_{GB}))d^5x+I_m$,
\end{center}
where $\alpha$, the coupling parameter has the dimension of $(length)^2$ and $I_m$ is the matter action.
Now, varying the action I over the metric tensor $g_{\mu \nu}$, we have the equations of motion: $G_{\mu \nu}-\alpha H_{\mu \nu}=T_{\mu \nu}$,
 where
\begin{equation}
 H_{\mu \nu}=4R_{\mu\lambda}R_{\nu}^\lambda+4R^{\rho \sigma}R_{\mu \rho \nu \sigma}-2RR_{\mu \nu}-2R_{\mu}^{\rho \sigma \lambda}R_{\nu \rho \sigma \lambda}+\frac{1}{2}g_{\mu \nu}R_{GB}
\end{equation}
is the Lovelock tensor, where $R_{GB}=R_{\mu \nu \rho \sigma}R^{\mu \nu \rho \sigma}-4R_{\mu \nu}R^{\mu \nu}+R^2 $.
Hence, the non-vanishing components of the modified Einstein's equations are
\begin{equation}\label{e1}
H^2 \left[1+\tilde{\alpha}H^2\right]=\frac{ \rho}{6}
\end{equation}
and
\begin{equation}\label{e2}
 [1+2\tilde{\alpha}H^2]\dot{H}=-\frac{1}{3}(\rho +p)
\end{equation}
\begin{figure}
\begin{minipage}{0.4\textwidth}
\includegraphics[width=1.0\linewidth]{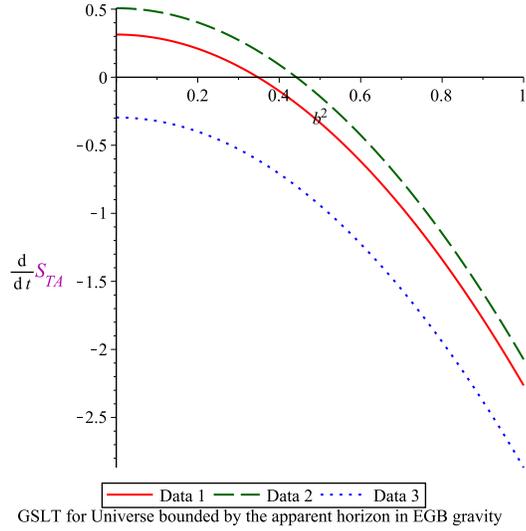}
(a)
\end{minipage}
\begin{minipage}{0.4\textwidth}
\includegraphics[width=1.0\linewidth]{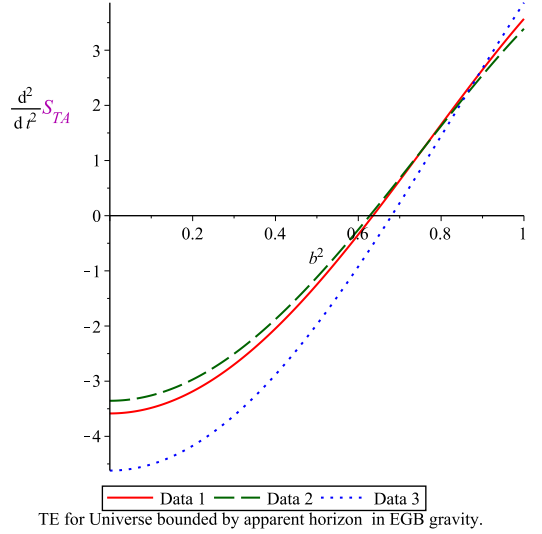}
(b)
\end{minipage}
\begin{minipage}{0.4\textwidth}
\includegraphics[width=1.0\linewidth]{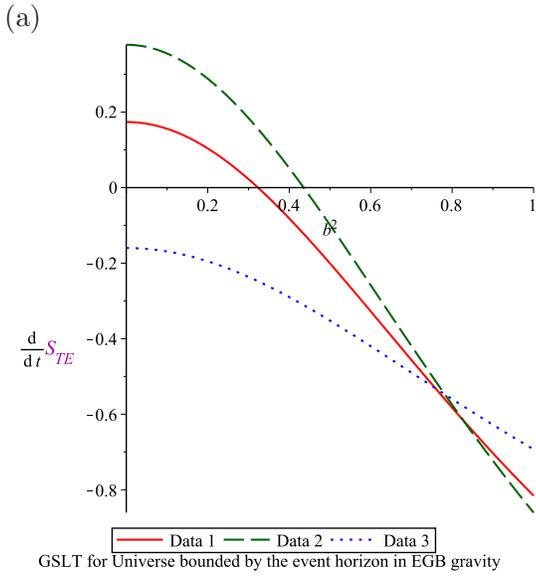}
(c)
\end{minipage}
\begin{minipage}{0.4\textwidth}
\includegraphics[width=1.0\linewidth]{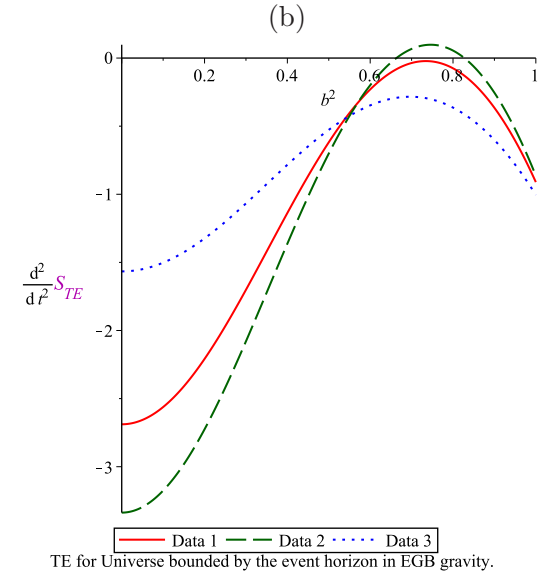}
(d)
\end{minipage}
\caption{The plots show GSLT and TE for Universe bounded by apparent/ event horizon for EGB gravity theory}
\end{figure}

Here $\tilde{\alpha}$ is the Gauss-Bonnet coupling parameter which is a function of $\alpha$.
Now from equations (\ref{e1}) and (\ref{e2}) we have
\begin{equation}
\rho_t=\rho-6\tilde{\alpha}H^4,
\end{equation}
\begin{equation}
 \rho_t+p_t=\rho +p+ 6\tilde{\alpha}\dot{H}H^2,
\end{equation}
\begin{equation}
\frac{\partial (\rho_e+p_e)}{\partial t}=-6\tilde{\alpha}H^4(f_A-4Hv_A^2)
\end{equation}

Assuming the validity of the unified first law, the expressions for entropy on the apparent and event horizon are obtained for EGB gravity  as:

\begin{equation}
 S_A=\frac{A_A}{4}+\frac{3}{4}\tilde{\alpha}ln(R_A),
\end{equation}
and
\begin{equation}
 S_E=\frac{A_E}{4}+\frac{3 \tilde{\alpha}}{4}\int \left(\frac{\epsilon}{1-\epsilon}\right)\frac{R_E}{R_A^2}\left(\frac{HR_E+1}{HR_E-1}\right)dR_E,
\end{equation}
Similar logarithmic correction to entropy is also obtained in black hole entropy of Einstein gravity with Gauss-Bonnet term \cite{a4}. Also it has been shown in Ref. \cite{a4} that even without log term in $S_A$, just for corresponding choice of EGB parameter in 5D EGB gravity, the corresponding entropy may become negative. Further, it is well-known that phantom cosmology which may be induced effectively by modified gravity, may also bring the negative entropy as well. Of course, the negative entropy often means the instability or maximum of corresponding action functional and as a result the study of thermodynamics at the apparent horizon will be difficult to analyze.

Now adding the fluid entropy with the horizon entropy, we have the first and second derivative of total entropy as,
\begin{equation}
\dot{S}_{TA}=\frac{v_A}{4}\left[R_A+3\tilde{\alpha}H+\frac{3}{16}R_A^2\frac{v_A-1}{2-v_A}\right],
\end{equation}
\begin{equation}
\dot{S}_{TE}=\frac{R_E v_E}{4}+\frac{3\tilde{\alpha}}{4}\left[\frac{v_AR_EH^2(v_E+2)}{2-v_A}\right]-\frac{3}{64}\frac{R_E^2 v_A}{2-v_A},
\end{equation}
\begin{equation}
\ddot{S}_{TA}=\frac{f_A}{4}\left[R_A+3\tilde{\alpha}H+\frac{3}{16}\left(\frac{R_A^2v_A(v_A-1)}{2-v_A}\right)\left(\frac{2v_A-1}{v_A(v_A-1)}+1\right)\right]+\frac{v_A^2}{4}\left[1-3\tilde{\alpha}H^2+\frac{3}{8}R_A\frac{v_A-1}{2-v_A}\right],
\end{equation}

\begin{equation}
\ddot{S}_{TE}=\frac{R_Ef_E}{4}\left(1+\frac{3\tilde{\alpha}}{4}\frac{v_AH^2}{2-v_A}\right)+\frac{v_E^2}{4}+\frac{3\tilde{\alpha}R_Ev_A}{4(2-v_A)}\{H^2(v_E+2)\left(\frac{2f_A}{v_A(2-v_A)}+\frac{v_E}{R_E}-2v_AH\right)-\frac{R_E}{8\tilde{\alpha}}\left(\frac{v_E}{R_E}+\frac{f_A}{v_A(2-v_A)}\right)\}.
\end{equation}

As before, $\dot{S}_{TA}, \ddot{S}_{TA}, \dot{S}_{TE}, \ddot{S}_{TE}$ have been plotted against $b^2$ for the  three Planck data sets \cite{r35} in Fig. 2(a)-(d) (considering $H=1, \tilde{\alpha}=2$) and analytic restrictions are presented in Table-IV.\\\\

\newpage
\begin{center}
{\bf Table IV}: Condition(s) required for GSLT and TE to hold in EGB-gravity
\end{center}
\begin{center}
\begin{tabular}{|p{2cm}|p{2cm}|p{13cm}|}
\hline \begin{center} GSLT/TE \end{center} & \begin{center} Horizon \end{center} & \begin{center} Condition(s) \end{center}\\
\hline \hline \begin{center} GSLT \end{center} & \begin{center} AH \end{center} &  \begin{center} $v_A(R_A+3\tilde{\alpha}H)-\frac{3}{16}R_A^2\frac{v_A(v_A-1)}{2-v_A}\geq 0$ \end{center}\\
\hline \begin{center} GSLT \end{center} & \begin{center} EH \end{center} &  \begin{center} $v_E \lesseqgtr \frac{3v_A(R_E-32\tilde{\alpha}H^2)}{16(2-v_A(1-3\tilde{\alpha}H^2))}$\end{center} \begin{center} according as \end{center}\begin{center} $\frac{\tilde{\alpha}v_AH^2}{2-v_A}\lessgtr \frac{-1}{3}$\end{center}\\
\hline \begin{center} TE \end{center} & \begin{center} AH \end{center} &  \begin{center} $f_A\left[R_A+3\tilde{\alpha}H+\frac{3}{16}\left(\frac{R_A^2v_A(v_A-1)}{2-v_A}\right)\left(\frac{2v_A-1}{v_A(v_A-1)}+1\right)\right]<-v_A^2\left[1-3\tilde{\alpha}H^2+\frac{3}{8}R_A\frac{v_A-1}{2-v_A}\right]$ \end{center}\\
\hline \begin{center} TE \end{center} & \begin{center} EH \end{center} & \begin{center} $\frac{R_Ef_E}{4}\left(1+\frac{3\tilde{\alpha}}{4}\frac{v_AH^2}{2-v_A}\right)<-\frac{v_E^2}{4}-\frac{3\tilde{\alpha}R_Ev_A}{4(2-v_A)}\{H^2(v_E+2)\left(\frac{2f_A}{v_A(2-v_A)}+\frac{v_E}{R_E}-2v_AH\right)-\frac{R_E}{8\tilde{\alpha}}\left(\frac{v_E}{R_E}+\frac{f_A}{v_A(2-v_A)}\right)\}$,\end{center} \\
\hline
\end{tabular}
\end{center}

\subsection{RSII brane world}

Brane world scenario is based on the assumption that our Universe is a 3-brane embedded in higher dimensional bulk space-time. Due to the extra dimension, the effective gravity on the brane is different from the standard Einstein gravity. The effective equations of motion on the 3-brane is
\begin{center}
$G_{\mu \nu}=-\Lambda_4q_{\mu \nu}+ G_4\tau_{\mu \nu}+\kappa_5^4\pi_{\mu \nu}-E_{\mu \nu}$
\end{center}
where\\
$$G_4=\frac{1}{6}\lambda \kappa_5^4,$$\\ $$\Lambda_4=\frac{\kappa_5^2}{2}\left(\Lambda_5+\frac{\kappa_5^2 \lambda^2}{6}\right),$$\\$$\pi_{\mu \nu}=\frac{-\tau_{\mu \alpha}\tau_{\nu}^\alpha}{4}+\frac{\tau \tau_{\mu \nu}}{12}+\frac{q_{\mu \nu}\tau_{\alpha \beta}\tau^{\alpha \beta}}{8}-\frac{q_{\mu \nu}\tau^2}{24},$$\\
and $E_{\mu \nu}$ is the electric part of the 5 dimensional Weyl tensor. Here $\lambda$ and $\tau_{\mu \nu}$ are the vaccum energy and energy-momentum tensor of matter on the brane, $\kappa_5, \Lambda_5~ and~ \Lambda_4$ are 5 dimensional gravity coupling constant, cosmological constant in the bulk and effective cosmological constant respectively.

The Friedmann equations in flat, homogeneous and isotropic brane,( without dark radiation term ) are given by \cite{r20,r36},
\begin{eqnarray}
 H^2&=&\frac{\rho_t}{3},
\nonumber
\\
\dot{H}&=&-\frac{1}{2}(\rho_t+p_t),
\end{eqnarray}
where  $$p_t=p+p_e$$
\begin{equation}
 \textrm{and}~~\rho_t=\rho+\frac{\kappa_5^4 \rho^2}{12}.
\end{equation}
The effective energy density $\rho_e$ and the effective pressure $p_e$, due to curvature contribution, are related by

\begin{equation}
 \rho_e+p_e=\frac{\kappa_5^4 \rho^2v_A}{9},
\end{equation}
differentiating which, we get
\begin{equation}
\frac{\partial (\rho_e+p_e)}{\partial t}=\frac{\kappa_5^4\rho^2}{9}(f_A-4Hv_A^2)
\end{equation}
As the entropy expressions are obtained using equations (\ref{28}) and (\ref{30}), so from the validity of the unified first law the expressions for entropy are obtained for RSII brane  as,
\begin{equation}
 S_A=\frac{A_A}{4}-\frac{ \kappa_5^4}{96}\int \frac{R_A^3}{\epsilon}\rho (\rho +p)dR_A.
\end{equation}
and

\begin{equation}
 S_E=\frac{A_E}{4}-\frac{ \kappa_5^4}{96}\int \left(\frac{R_A^2 R_E}{1-\epsilon}\right)\left(\frac{HR_E+1}{HR_E-1}\right)\rho (\rho+p)dR_E.
\end{equation}
Now, adding the fluid entropy with the horizon entropy for apparent and event horizon respectively, we have the first and second derivatives of the total entropy as,
\begin{equation}
\dot{S}_{TA}=\frac{R_A v_A}{4}-\frac{R_A^3}{8}(\rho_e+p_e)+\frac{R_A^3v_A(v_A-1)\rho_t}{6(2-v_A)},
\end{equation}
\begin{equation}
\dot{S}_{TE}=\frac{R_Ev_E}{4}-\frac{R_A^2R_E}{2(2-v_A)}\{\frac{v_E+2}{4}(\rho_e+p_e)+\frac{v_A\rho_t}{3}\},
\end{equation}
\begin{equation}
\ddot{S}_{TA}=\frac{R_Af_A}{4}\{1-\frac{2R_A^2\rho_t}{3(v_A-2)^2}(v_A^2-4v_A+2)\}-\frac{R_A^2}{8}\{3v_A(\rho_e+p_e)+8R_A\frac{\partial (\rho_e+p_e)}{\partial t}\}-\frac{R_A^2v_A^2(v_A-1)\rho_t}{6(v_A-2)},
\end{equation}
$$\ddot{S}_{TE}=\frac{R_Ef_E}{4}\{1-\frac{R_A^2(\rho_e+p_e)}{2(2-v_A)}\}+\frac{v_E^2}{4}-\frac{R_A^2R_E}{8(2-v_A)}\{(v_E+2)\left(2\frac{v_A}{R_A}+\frac{v_E}{R_E}+\frac{f_A}{2-v_A}\right)(\rho_e+p_e)$$
\begin{equation}
+\frac{\partial (\rho_e+p_e)}{\partial t}\}+\frac{4v_A\rho_t}{3}\{\frac{v_E}{R_E}+\frac{2f_A}{v_A(2-v_A)}\}.
\end{equation}
Here again, $\rho_t$, $(\rho_e+p_e)$, $\frac{\partial (\rho_e+p_e)}{\partial t}$ are to be substituted from equations (67)-(69).
$\dot{S}_{TA}, \ddot{S}_{TA}, \dot{S}_{TE}, \ddot{S}_{TE}$ have been plotted against $b^2$ for the three Planck data sets as before  in Fig. 3 (a)-(d) (considering $\kappa_5=1, H=1 $). The inequalities corresponding to (\ref{1}) are shown in Table-V.

\begin{figure}
\begin{minipage}{0.4\textwidth}
\includegraphics[width=1.0\linewidth]{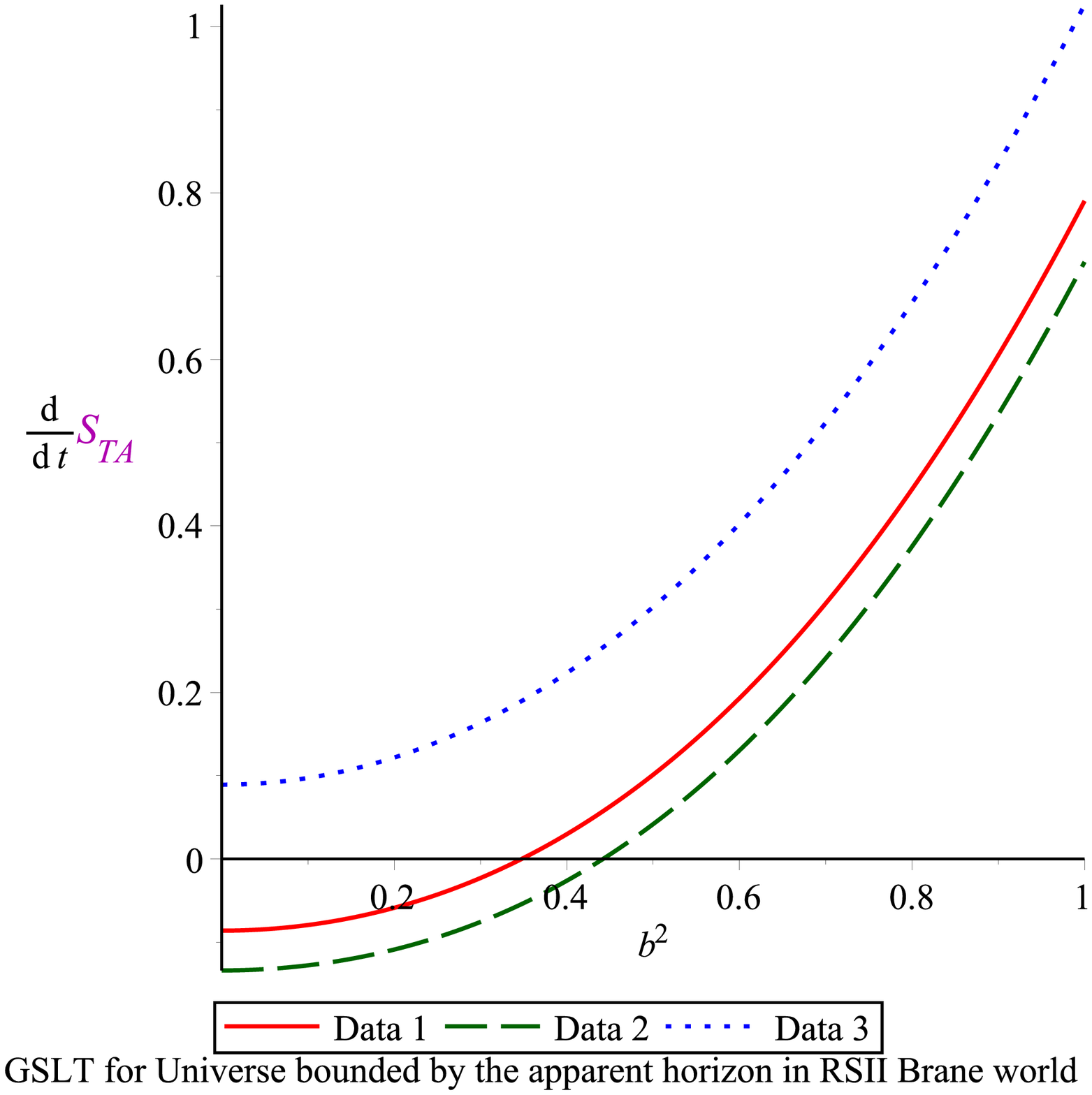}
(a)
\end{minipage}
\begin{minipage}{0.4\textwidth}
\includegraphics[width=1.0\linewidth]{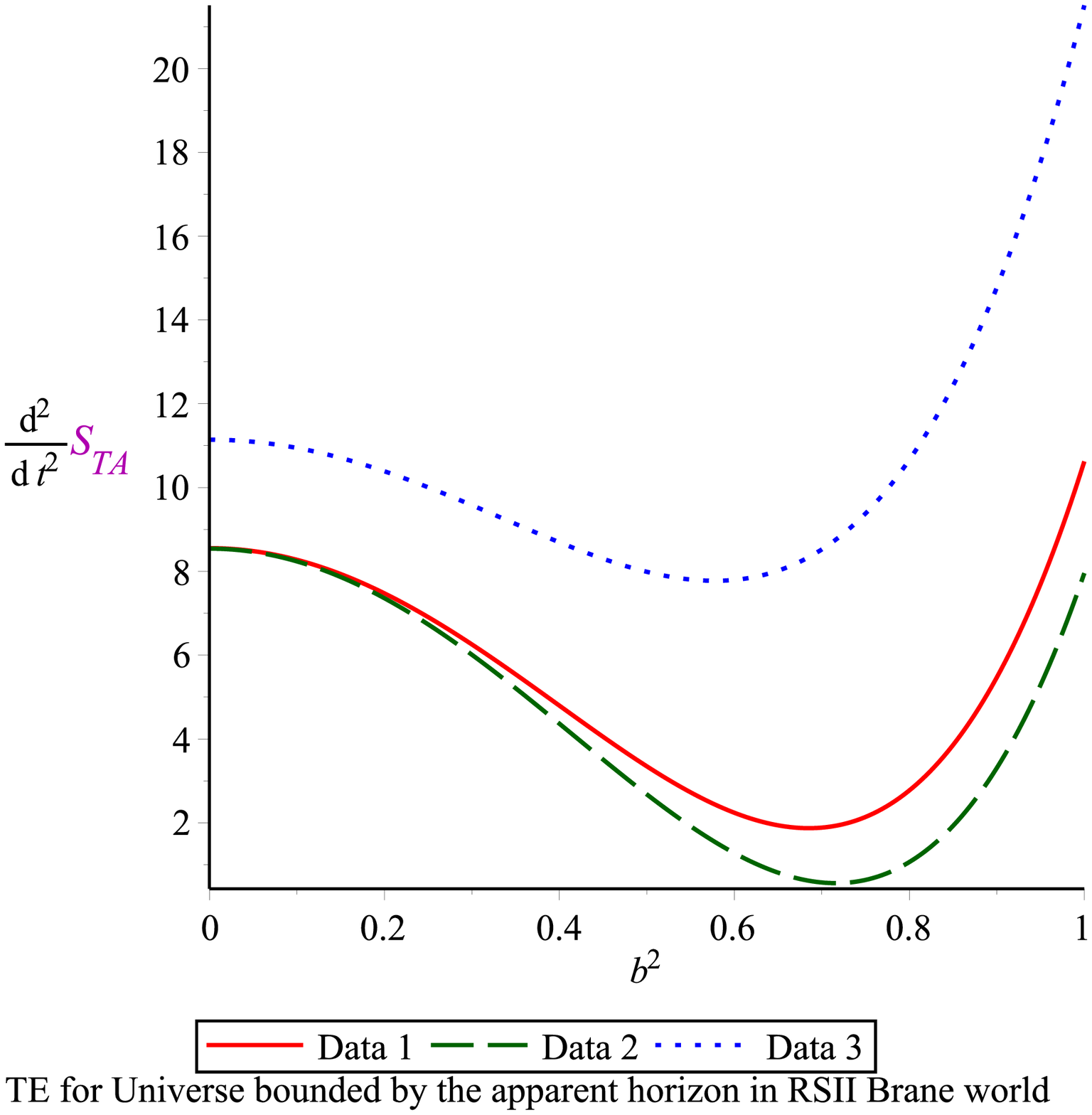}
(b)
\end{minipage}
\begin{minipage}{0.4\textwidth}
\includegraphics[width=1.0\linewidth]{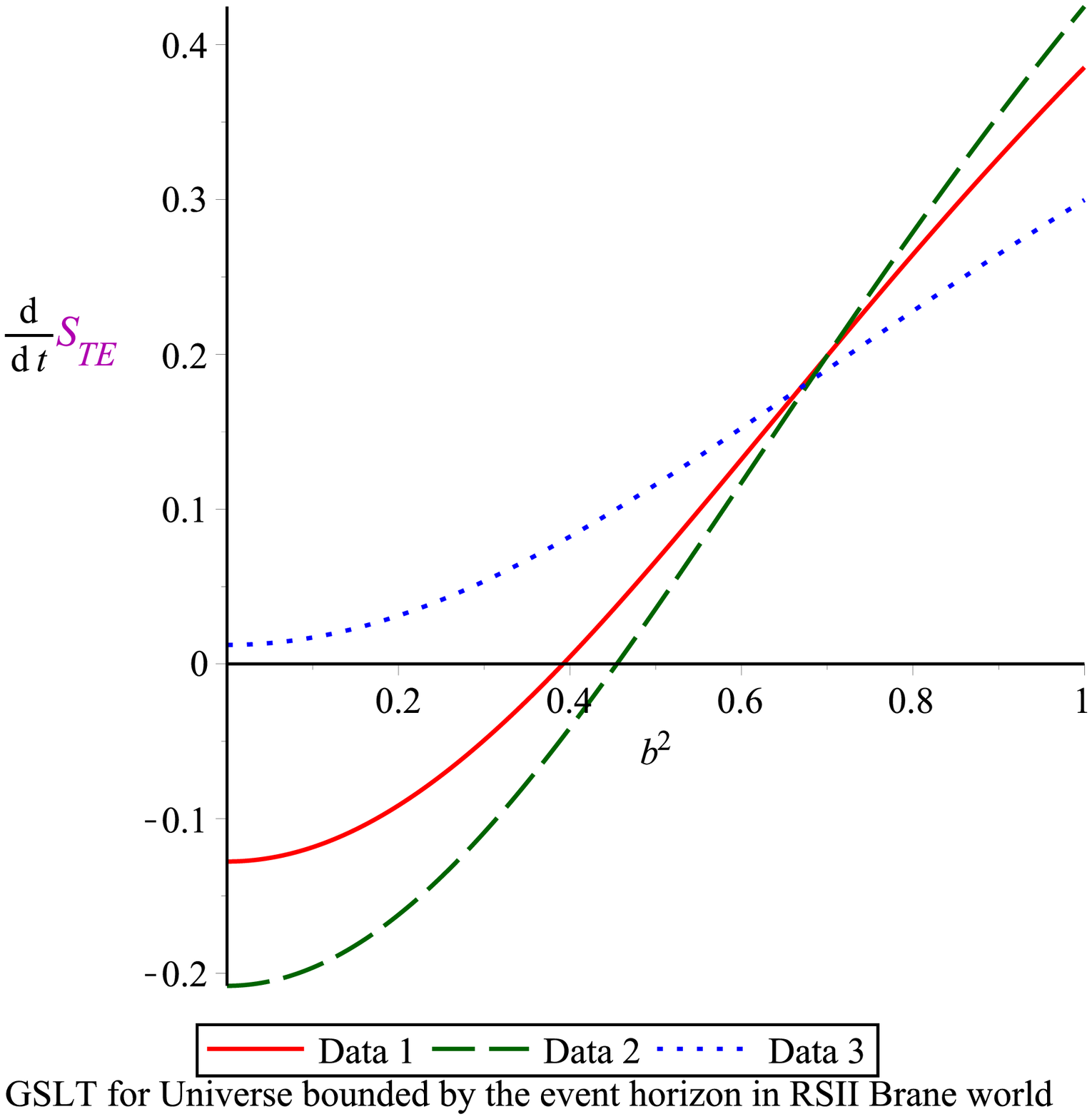}
(c)
\end{minipage}
\begin{minipage}{0.4\textwidth}
\includegraphics[width=1.0\linewidth]{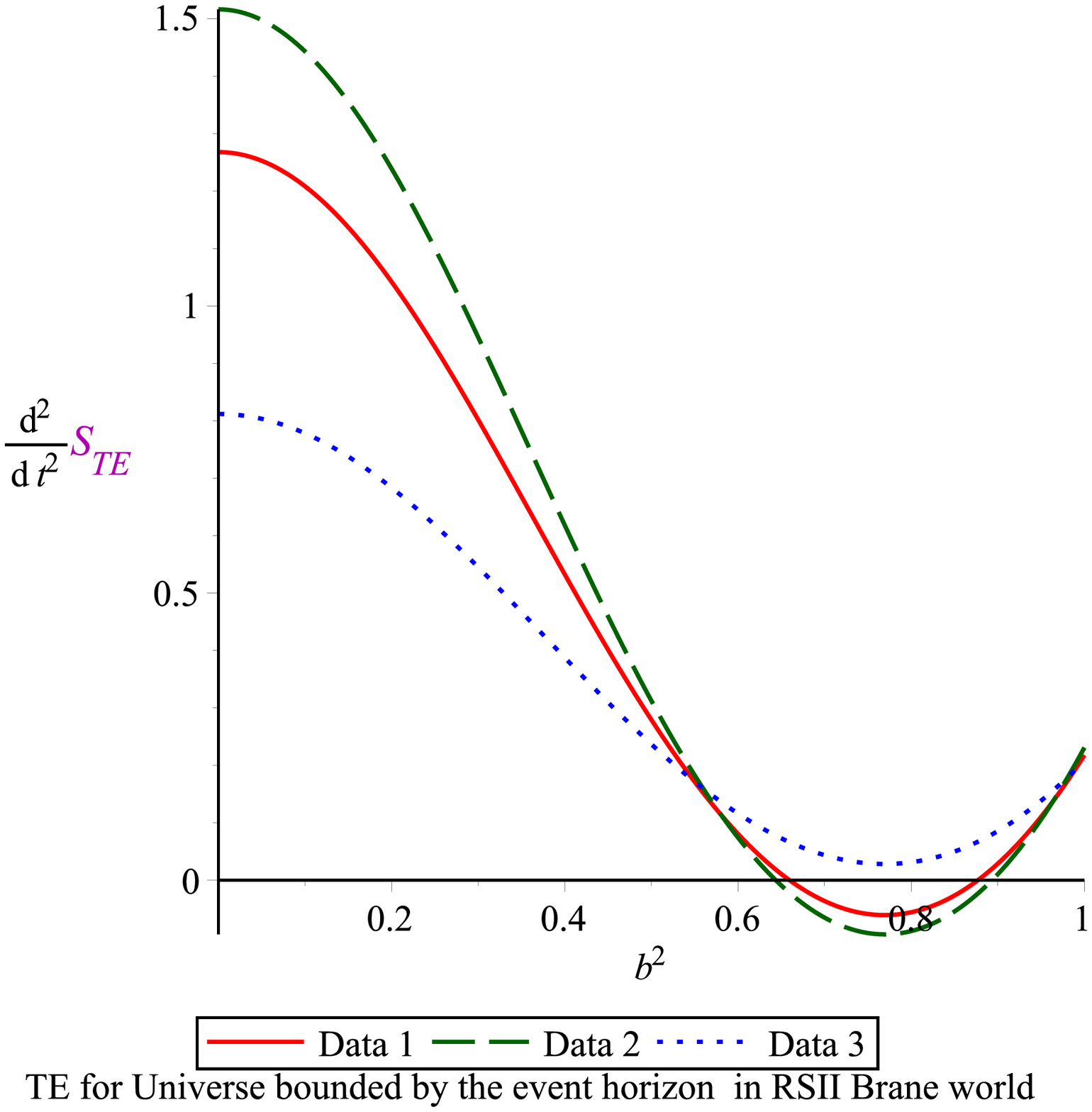}
(d)
\end{minipage}
\caption{The plots show GSLT and TE for Universe bounded by apparent/ event horizon for RSII brane model}
\end{figure}

\newpage
\begin{center}
{\bf Table V}: Condition(s) required for GSLT and TE to hold in RSII brane
\end{center}
\begin{center}
\begin{tabular}{|p{1.6cm}|p{1.2cm}|p{15cm}|}
\hline \begin{center} GSLT/TE \end{center} & \begin{center} Horizon \end{center} & \begin{center} Condition(s) \end{center}\\
\hline \hline \begin{center} GSLT \end{center} & \begin{center} AH \end{center} &  \begin{center} $\frac{2R_A^2\rho (v_A-1)(1-\frac{\kappa_5^4\rho}{12})}{3(v_A-2)}+\frac{R_A^2\kappa_5^4\rho^2}{18}-1 \lesseqgtr 0$ \end{center} \begin{center} according as $v_A \gtrless 0$ \end{center} \\
\hline \begin{center} GSLT \end{center} & \begin{center} EH \end{center} &  \begin{center} $v_E \lesseqgtr \frac{2R_A^2v_A \rho\{1+\frac{\kappa_5^4 \rho}{12}\}}{3\{2-v_A\left(1+\frac{R_A^2\kappa_5^4\rho^2}{18}\right)\}}$\end{center} \begin{center}
according as $1\lessgtr \frac{R_A^2\kappa_5^4\rho^2v_A}{18(2-v_A)}$ \end{center}\\
\hline \begin{center} TE \end{center} & \begin{center} AH \end{center} &  \begin{center} $f_A \left[1-\frac{2R_A^2\rho(1-\frac{\kappa_5^4\rho}{12})(v_A^2-4v_A+2)}{3(v_A-2)^2}-\frac{4}{9}R_A^2\kappa_5^4\rho^2\right]<\frac{R_Av_A^2}{6}\left[\frac{-29}{3}\kappa_5^4\rho^2+\frac{4R_A\rho (v_A-1)(1-\frac{\kappa_5^4\rho}{12})}{(v_A-2)}\right]$ \end{center}\\
\hline \begin{center} TE \end{center} & \begin{center} EH \end{center} & \begin{center} $f_E \left[2(2-v_A)-\frac{R_A^2\kappa_5^4\rho^2v_A}{9}\right]\lessgtr R_A^2\left[(v_E+2)\frac{\kappa_5^4\rho^2}{9}\{v_A(2\frac{v_A}{R_A}+\frac{v_E}{R_E}+\frac{f_A}{2-v_A})+(f_A-4\frac{v_A^2}{R_A})\}\right.$\end{center}\begin{center}$\left.+\frac{4}{3}v_A\rho(1-\frac{\kappa_5^4\rho}{12})(\frac{v_E}{R_E}+\frac{2f_A}{v_A(2-v_A)})\right]-2\frac{v_E^2}{R_E}(2-v_A)$ \end{center} \begin{center}
according as $v_A\lessgtr 2$
\end{center}\\
\hline
\end{tabular}
\end{center}

\subsection{DGP brane world}

DGP brane world model is a simple and well studied model of brane-gravity. In this model, our four-dimensional world is a FRW brane embedded in a five-dimensional Minkowski bulk. The action of gravity is proportional to $M_P^2$ ($M_P$ is the Planck mass in four dimension) in the four-dimensional brane, whereas in the bulk it is proportional to the corresponding quantity in five dimensions. The model is then characterized by a crossover length scale
\begin{center}
$r_c=\frac{M_P^2}{2M_5^2}$
\end{center}
such that gravity is a four-dimensional theory at scales $a\ll r_c$ where matter behaves as pressure less dust but gravity leaks out into the bulk at scales $a\gg r_c$ and matter approaches the behavior of a cosmological constant.
In a flat, homogeneous and isotropic brane, the Friedmann equation in DGP model is given by (\cite{b1,r37})
\begin{equation}
 H^2-\tilde{\epsilon}\frac{H}{r_c}=\frac{\rho}{3}
\end{equation}
where $\tilde{\epsilon}=\pm1$ corresponds to standard DGP(+) model (self accelerating without any form of dark energy) and DGP(-) model (not self accelerating, requires dark energy)
respectively.

From equation (76) and using the conservation equation $\dot{\rho}+3H(\rho +p)=0$, it can be shown that
\begin{equation}
 \dot{H}=-\frac{1}{2}\left[\rho+p +\frac{\tilde{\epsilon} (\rho +p)}{2Hr_c-\tilde{\epsilon}}\right].
\end{equation}
Hence,
\begin{equation}
\rho_t=\rho+\frac{3\tilde{\epsilon}H}{r_c},
\end{equation}
\begin{figure}
\begin{minipage}{0.4\textwidth}
\includegraphics[width=1.0\linewidth]{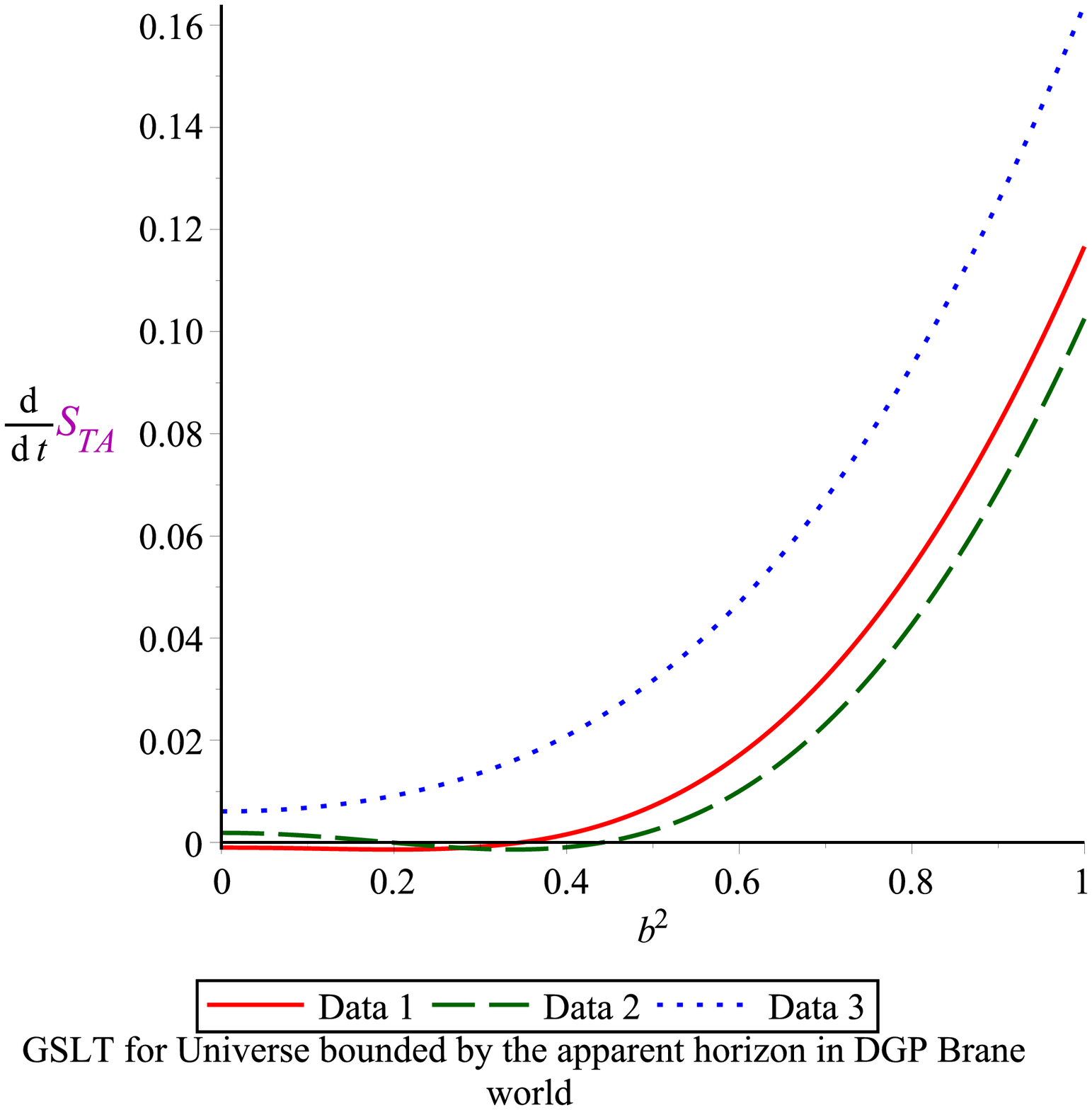}
(a)
\end{minipage}
\begin{minipage}{0.4\textwidth}
\includegraphics[width=1.0\linewidth]{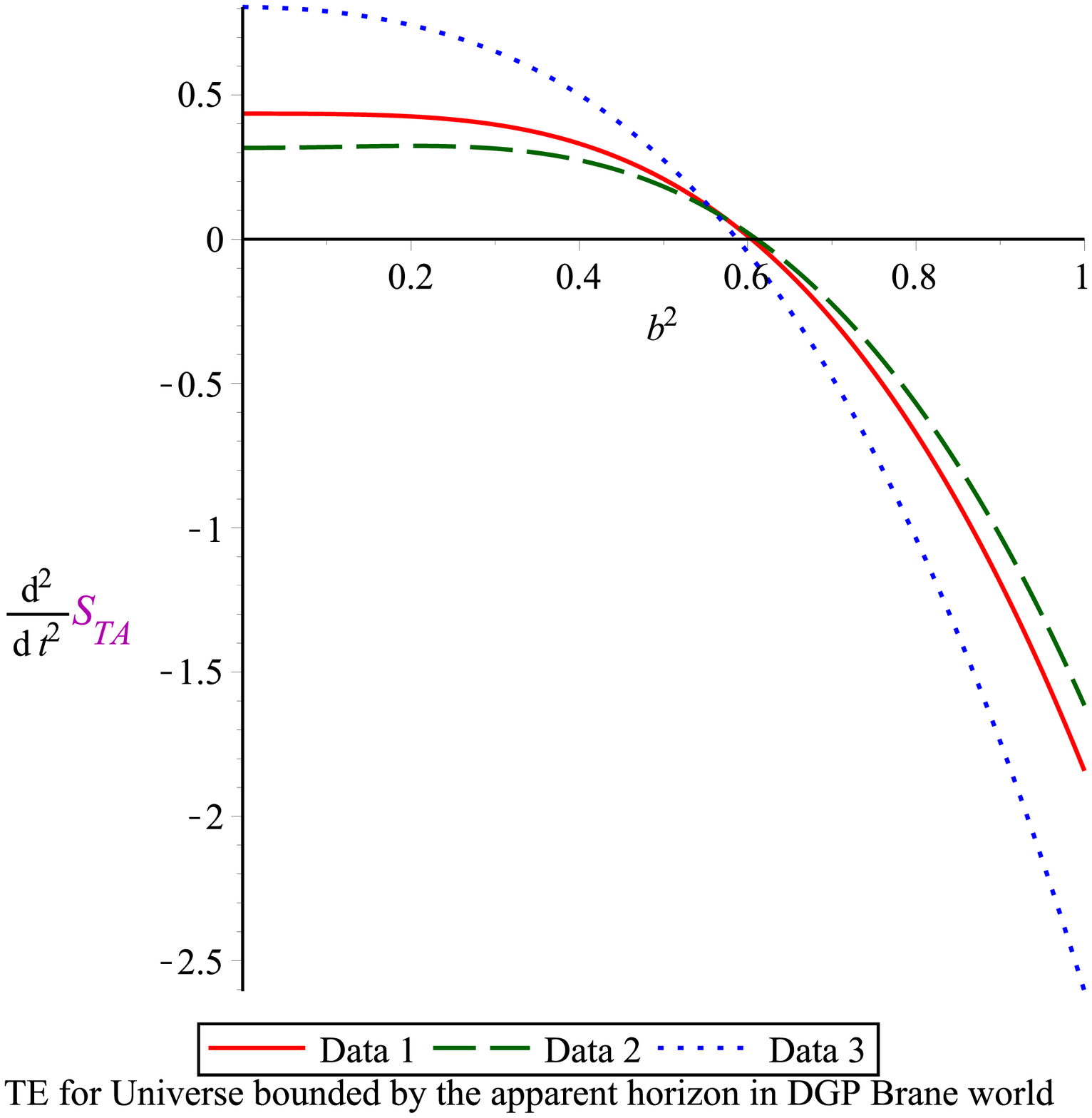}
(b)
\end{minipage}
\begin{minipage}{0.4\textwidth}
\includegraphics[width=1.0\linewidth]{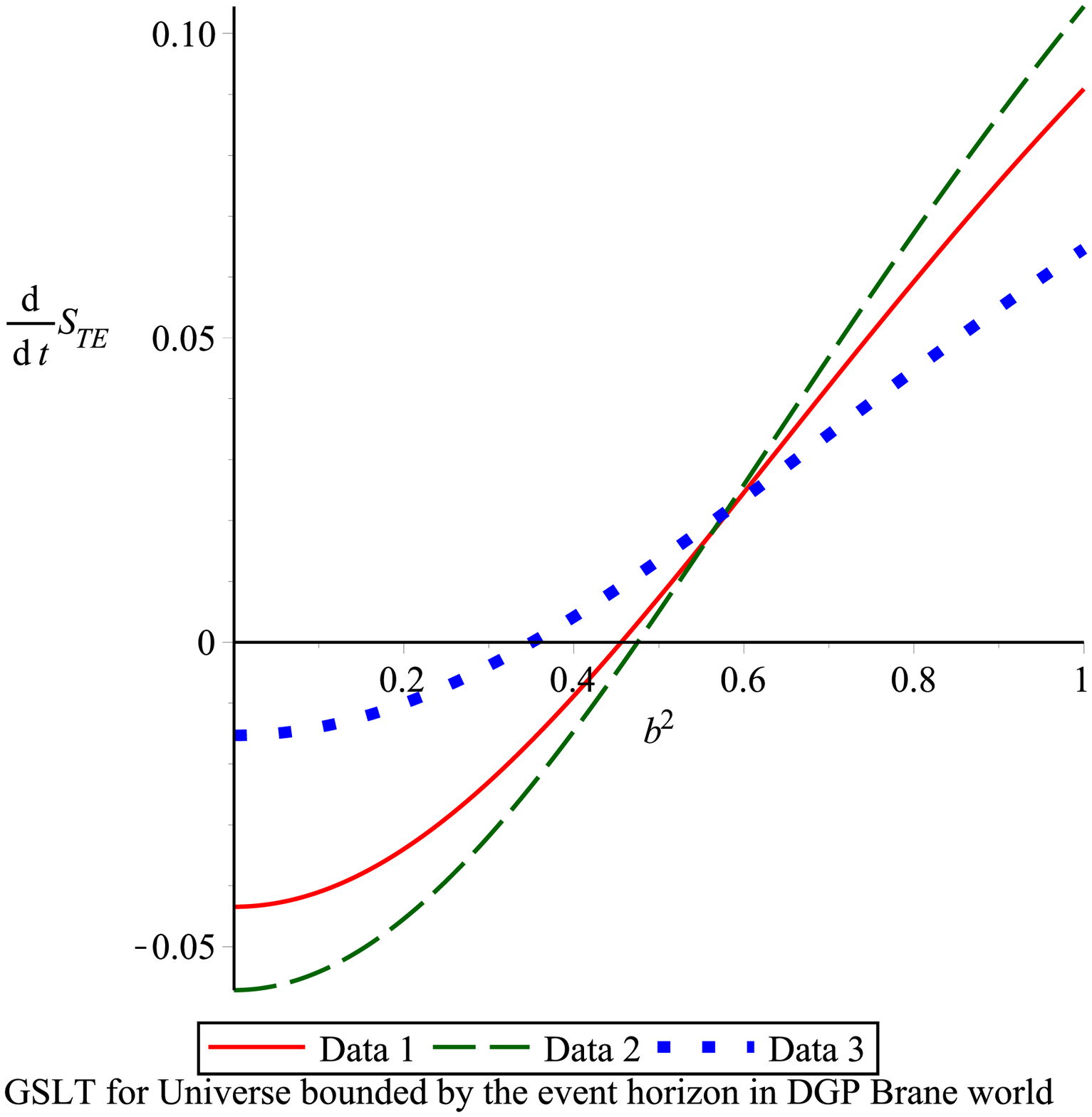}
(c)
\end{minipage}
\begin{minipage}{0.4\textwidth}
\includegraphics[width=1.0\linewidth]{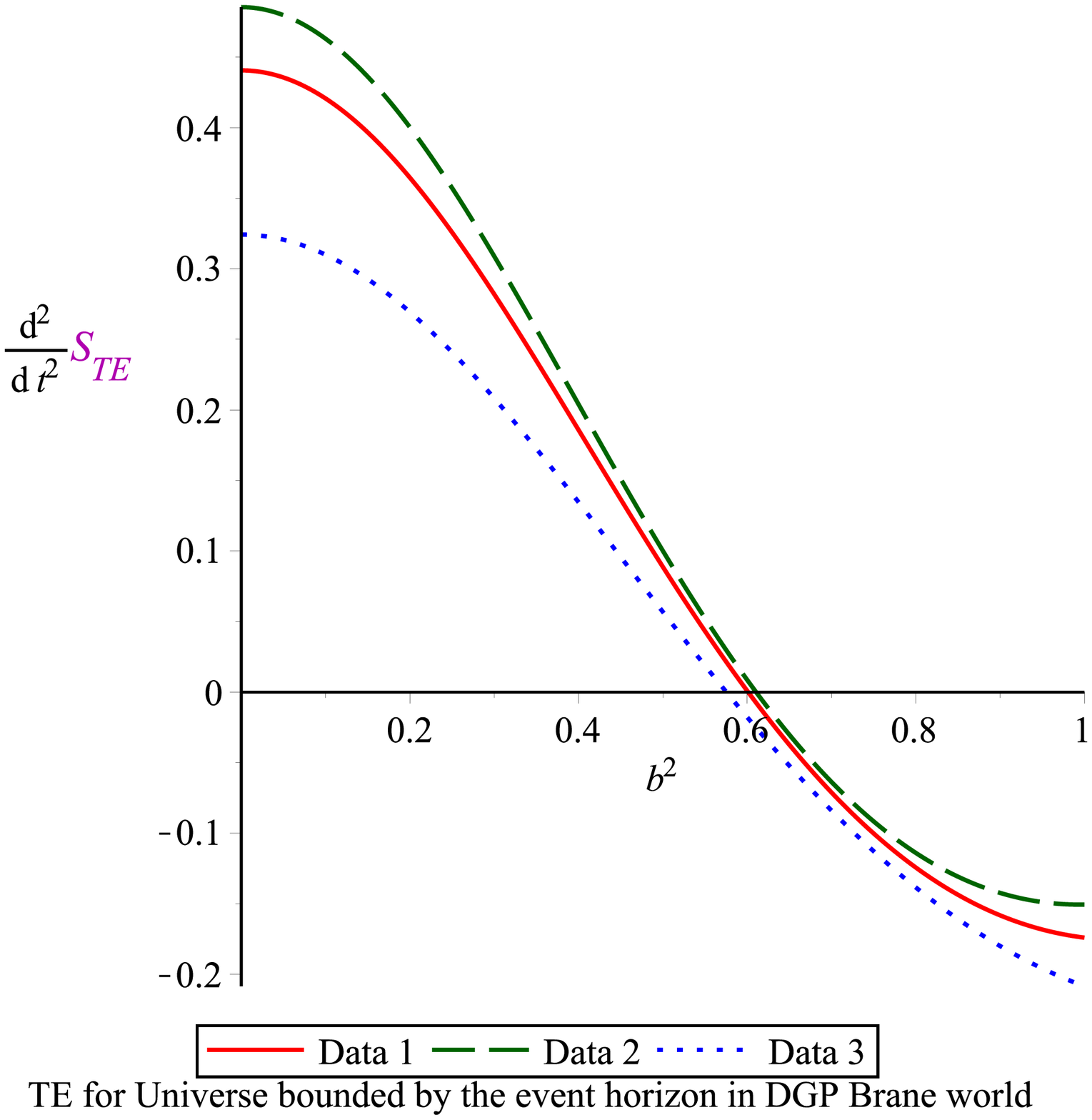}
(d)
\end{minipage}
\caption{The plots show GSLT and TE for Universe bounded by apparent/ event horizon for DGP brane model}
\end{figure}

\begin{equation}
 \rho_e+p_e=\frac{2\tilde{\epsilon}\rho v_A}{3(2Hr_c-\tilde{\epsilon})}.
\end{equation}
and
\begin{equation}
\frac{\partial (\rho_e+p_e)}{\partial t}=\frac{2\rho \tilde{\epsilon}}{3(2Hr_c-\tilde{\epsilon})}\left(f_A-2Hv_A^2+\frac{2H^2v_A^2r_c}{2Hr_c-\tilde{\epsilon}}\right)
\end{equation}

Assuming the validity of the unified first law, the expressions for entropy on the apparent and event horizon are obtained for DGP brane  using equations (\ref{28}) and (\ref{30}) ,
\begin{equation}
 S_A=\frac{A_A}{4}-\frac{1}{16} \tilde{\epsilon} \int \left(\frac{R_A^3}{\epsilon}\right)\left(\frac{\rho+p}{2Hr_c-\tilde{\epsilon}}\right)dR_A,
\end{equation}
and
\begin{equation}
 S_E=\frac{A_E}{4G}-\frac{1}{16}\tilde{\epsilon}\int \left(\frac{R_A^2 R_E}{1-\epsilon}\right)\left(\frac{HR_E+1}{HR_E-1}\right)\frac{\rho+p}{2Hr_c-\tilde{\epsilon}}dR_E,
\end{equation}
Now, respectively adding the fluid entropy with the horizon entropy, we have the first and second derivatives of total entropy as,
\begin{equation}
\dot{S}_{TA}=\frac{R_A v_A}{4}-\frac{R_A^3}{8}(\rho_e+p_e)+\frac{R_A^3v_A(v_A-1)\rho_t}{6(2-v_A)},
\end{equation}
\begin{equation}
\dot{S}_{TE}=\frac{R_Ev_E}{4}-\frac{R_A^2R_E}{2(2-v_A)}\{\frac{v_E+2}{4}(\rho_e+p_e)+\frac{v_A\rho_t}{3}\},
\end{equation}
\begin{equation}
\ddot{S}_{TA}=\frac{R_Af_A}{4}\{1-\frac{2R_A^2\rho_t}{3(v_A-2)^2}(v_A^2-4v_A+2)\}-\frac{R_A^2}{8}\{3v_A(\rho_e+p_e)+8R_A\frac{\partial (\rho_e+p_e)}{\partial t}\}-\frac{R_A^2v_A^2(v_A-1)\rho_t}{6(v_A-2)},
\end{equation}
$$\ddot{S}_{TE}=\frac{R_Ef_E}{4}\{1-\frac{R_A^2(\rho_e+p_e)}{2(2-v_A)}\}+\frac{v_E^2}{4}-\frac{R_A^2R_E}{8(2-v_A)}\{(v_E+2)\left(2\frac{v_A}{R_A}+\frac{v_E}{R_E}+\frac{f_A}{2-v_A}\right)(\rho_e+p_e)$$
\begin{equation}
+\frac{\partial (\rho_e+p_e)}{\partial t}\}+\frac{4v_A\rho_t}{3}\{\frac{v_E}{R_E}+\frac{2f_A}{v_A(2-v_A)}\}.
\end{equation}
Here, $\rho_t$, $(\rho_e+p_e)$, $\frac{\partial (\rho_e+p_e)}{\partial t}$ are to be substituted from equations (78)-(80).

$\dot{S}_{TA}, \ddot{S}_{TA}, \dot{S}_{TE}, \ddot{S}_{TE}$ have been plotted against $b^2$ for the three Planck data sets [35] in Fig. 4 (a)-(d) (considering $\tilde{\epsilon}=1, H=1.5, r_c=1$) and Table-VI shows the restrictions analytically.

\begin{center}
{\bf Table VI}: Condition(s) required for GSLT and TE to hold in DGP brane
\end{center}
\begin{center}
\begin{tabular}{|p{1.6cm}|p{1.2cm}|p{15cm}|}
\hline \begin{center} GSLT/TE \end{center} & \begin{center} Horizon \end{center} & \begin{center} Condition(s) \end{center}\\
\hline \hline \begin{center} GSLT \end{center} & \begin{center} AH \end{center} &  \begin{center} $\frac{2R_A^2(v_A-1)}{3(v_A-2)}(\rho +\frac{3\tilde{\epsilon}H}{r_c})+\frac{\rho R_A^2 \tilde{\epsilon}}{3(2Hr_c-\tilde{\epsilon})}-1\lesseqgtr 0$ \end{center} \begin{center} according as $v_A\lessgtr 0$ \end{center}\\
\hline \begin{center} GSLT \end{center} & \begin{center} EH \end{center} &  \begin{center} $v_E \lesseqgtr \frac{2v_A R_A \{ 2\rho r_c+3\frac{\tilde{\epsilon}}{r_c}(2Hr_c-\tilde{\epsilon})\}}{3(2-v_A)(2Hr_c-\tilde{\epsilon})-R_A^2\rho v_A\tilde{\epsilon}}$\end{center} \begin{center}
according as $1\lessgtr \frac{R_A^2\rho v_A}{3(2-v_A)(2Hr_c-\tilde{\epsilon})}$
\end{center}\\
\hline \begin{center} TE \end{center} & \begin{center} AH \end{center} &  \begin{center} $f_A \left[1-\frac{8\rho \tilde{\epsilon}R_A^2}{3(2Hr_c-\tilde{\epsilon})}-\frac{2R_A^2(v_A^2-4v_A+2)}{3(v_A-2)^2}(\rho+\frac{3\tilde{\epsilon}H}{r_c})\right]<\frac{R_Av_A^2}{3}\left[\frac{\rho \tilde{\epsilon}(6Hr_c-13\tilde{\epsilon})}{(2Hr_c-\tilde{\epsilon})^2}+\frac{2R_A(v_A-1)}{(v_A-2)}(\rho+\frac{3\tilde{\epsilon}H}{r_c})\right]$ \end{center}\\
\hline \begin{center} TE \end{center} & \begin{center} EH \end{center} & \begin{center} $f_E \left[(2-v_A)-\frac{\rho \tilde{\epsilon}R_A^2v_A}{3(2Hr_c-\tilde{\epsilon})}\right]\lessgtr R_A^2\left[(v_E+2)\frac{\rho \tilde{\epsilon}}{3(2Hr_c-\tilde{\epsilon})}\{v_A(2\frac{v_A}{R_A}+\frac{v_E}{R_E}+\frac{f_A}{2-v_A})+f_A-2Hv_A^2(1-\frac{Hr_c}{2Hr_c-\tilde{\epsilon}})\}\right.$\end{center}\begin{center}$\left.+\frac{4}{3}v_A(\frac{v_E}{R_E}+\frac{2f_A}{v_A(2-v_A)})(\rho+\frac{3\tilde{\epsilon}H}{r_c})\right]-\frac{v_E^2}{R_E}(2-v_A)$,\end{center} \begin{center} according as $v_A\lessgtr 2$ \end{center}\\
\hline
\end{tabular}
\end{center}

\section{Discussion}

In the present work, we have examined the validity of GSLT and TE for FRW Universe bounded by apparent/event horizon for different gravity theories. An ideal thermodynamical system should obey the generalized second law of thermodynamics and it should be in thermal equilibrium. From the point of view of present accelerating phase we have chosen HDE (with event horizon as IR cut off) interacting with DM as the matter contained in the Universe. Although we are considering purely classical ideal fluid, but from the point of view of thermal equilibrium with the horizon, we need a quantum description. This is possible if we consider the fluid under consideration as an effective kind of description of a real scalar field $\phi$ having self interacting potential $V(\phi)$. As the time variations of the total entropy (for both the horizons) is complicated so we can not make any conclusion from the restrictions (presented in tabular form) for the validity of GSLT and TE. As a result, using recent Planck data sets for estimation of parameters in HDE, we have presented the total entropy variations graphically.

From the figures 1(a)-(d) we see that GSLT does not hold for any data set on both the horizons in f(R)-gravity with $f(R)=R+R^2$. The thermodynamical equilibrium (TE) does not hold for the first two data sets (except for a small neighbourhood of $b^2=0.6$) for both the horizons, but we have some peculiar situation for the 3rd data set. Here TE at the apparent horizon holds for all $b^2$ while at the event horizon TE holds for $0.2 \lesssim b^2 \lesssim 0.8$. As in a thermodynamical system GSLT does not holds but TE holds, which is not a realistic situation, so we may say that the third data set is not appropriate in the present case and for the first two data sets Universe as a thermodynamical system is not an ideal one in f(R)-gravity model.

In Einstein-Gauss-Bonnet gravity, for the third data set, the thermodynamical system has a similar (as above in f(R)-gravity) contradictory behaviour and hence we discard this data set for the present EGB gravity theory. On the other hand, Universal thermodynamics for Universe bounded by apparent/event horizon in EGB gravity theory is an ideal thermodynamical system provided the coupling parameter '$b^2$' is restricted to (see figures 2(a)-(d)) $0 \lesssim b^2 \lesssim 0.4$ for both the horizons.

From figures 3(a)-(d), we observe that in RSII brane scenario, the situation is not so worse for the third data set-GSLT holds at both the horizons for all values of '$b^2$' but not the thermodynamical equilibrium. On the other hand, for the first two data sets GSLT holds for $0.4 \lesssim b^2$ at both the horizons but there is no longer any thermal equilibrium (except for a small range of '$b^2$' around $b^2=0.8$ at the event horizon). So Universe bounded by any of the two horizons in RS-II brane model is not an ideal thermodynamical system for all the three data sets.

Lastly, in DGP brane model we see from figures 4(a)-(d) that both GSLT and TE hold at both the horizons and for all the data sets provided $b^2$ is restricted to $0.6 \lesssim b^2 $. Hence Universal thermodynamics in DGP brane model for Universe bounded by apparent/ event horizon is a possible ideal thermodynamical system (with $0.6 \lesssim b^2 $).

Further, it should be mentioned that similar analysis is possible for scalar tensor theory of gravity. But due to very complicated expressions for the time derivatives of the total entropy, it is not possible to conclude any thing either graphically or from tabular representation. Hence we have not presented it here.

In the present work the modified gravity theories are represented (as usually done) in the form of generalized fluid with inhomogeneous equation of state. Usually, this type of theories may pretend to unify the early-time inflation with the theory describing the late-time acceleration. Also, it has been shown that phantom scalar field models can be mapped into a mathematically equivalent, modified f(R) gravity. Further, modified gravity becomes complex at the region where the original phantom dark energy theory develops a Big Rip singularity . Thus the thermodynamical analysis may be questionable in the regime near to one of the four types of future singularities (as classified in Ref. \cite{a8}). In this context, it should be noted that in Ref. \cite{a7} it has been shown that except for some special cases of Type II and Type IV singularities, the dynamical entropy bound is violated near the singularity. Hence, for future work  in the context of GSLT and TE, it will be interesting to study the validity of them and the dynamical entropy bound near the future singularities. Also it will be interesting to examine the dynamical entropy bound of the modified entropies derived in the present work for both apparent and event horizons.

Therefore, from the above discussion we conclude that similar to Einstein gravity \cite{r30} (where event horizon is more favourable than apparent horizon) we have no definite conclusion whether apparent or event horizon is more favourable from thermodynamic view point in different gravity theories i.e., both the horizons are equally favoured or disfavoured from thermodynamic view point in different gravity theories.

\section*{Acknowledgement}
The author S.M. is thankful to UGC for NET-JRF.
The author S.S. is thankful to UGC-BSR Programme of Jadavpur University for awarding Research Fellowship.
S.C. is thankful to UGC-DRS programme, Department of Mathematics, J.U.

\end{document}